\documentclass[preprint,prb,aps, amsmath,amssymb,showpacs,superscriptaddress,seceqn]{revtex4-1}
\usepackage{amsfonts, hyperref, times}
\usepackage{graphicx}
\usepackage{siunitx}

\providecommand{\vect}[1]{{\boldsymbol{#1}}}
\DeclareMathOperator{\sgn}{sign}

\begin{document}

\title{Skyrmion production on demand by homogeneous DC currents}

\author{Karin Everschor-Sitte}
\affiliation{Institute of Physics, Johannes Gutenberg-Universit{\"a}t, 55128 Mainz, Germany}

\author{Matthias Sitte}
\affiliation{Institute of Physics, Johannes Gutenberg-Universit{\"a}t, 55128 Mainz, Germany}

\author{Thierry Valet}
\affiliation{Institute of Physics, Johannes Gutenberg-Universit{\"a}t, 55128 Mainz, Germany}

\author{Ar.~Abanov}
\affiliation{Department of Physics \& Astronomy, Texas A\&M University, College Station, Texas 77843-4242, USA}

\author{Jairo Sinova}
\affiliation{Institute of Physics, Johannes Gutenberg-Universit{\"a}t, 55128 Mainz, Germany}
\affiliation{Institute of Physics, Academy of Sciences of the Czech Republic, Cukrovarnicka 10, 162 53 Praha 6 Czech Republic}

\date{\today}

\begin{abstract}
Topological magnetic textures -- like skyrmions -- have become a major player in the design of next-generation magnetic storage technology due to their stability and the control of their motion by ultra-low current densities.
A major challenge to develop this new skyrmion-based technology is to achieve the controlled and deterministic creation of magnetic skyrmions without the need of complex setups.
We demonstrate a solution to this challenge by showing how to create skyrmions and other magnetic textures in ferromagnetic thin films by means of a homogeneous DC current and without requiring Dzyaloshinskii-Moriya interactions.
This is possible by exploiting a static loss of stability arising from the interplay of current-induced spin-transfer torque and a spatially inhomogeneous magnetization, which can be achieved, \textit{e.g.}, by locally engineering the anisotropy, the magnetic field, or other magnetic interactions.
The magnetic textures are created controllably, efficiently, and periodically with a period that can be tuned by the applied current strength.
We propose specific experimental setups realizable with simple materials, such as cobalt based materials, to observe the periodic formation of skyrmions.
We show that adding chiral interactions will not influence the basics of the generations but then influence the consequent dynamics with respect to the stabilization of topological textures.
Our findings allow for the production of skyrmions on demand in simple ferromagnetic thin films by homogeneous DC currents.
\end{abstract}


\maketitle

\section{Introduction}

Technologies based on spintronics have become integral parts of our world.
Current mass-market magnetic memory technologies primarily rely on spintronic devices that couple to the magnetic fields created by domains, which brings inherent limits in storage density and speed.
The next-generation, high-performance magnetic memory devices rely on the ability to efficiently and controllably create and manipulate magnetic textures by purely electrical means.
Magnetic storage devices based on the racetrack memory idea, where traditionally the information is encoded via magnetic domain walls, has been proposed as a design of ultra-dense, low-cost and low-power storage technologies.\cite{Parkin2008}
However, several difficulties arise to efficiently control the domain walls:
i) large current densities are needed to move them;
ii) nanowires of high quality are required as edge roughnesses will modify the shape of the domain walls or even destroy them; and
iii) the spacing between two magnetic domains can hardly be reduced below 30-40 nm.\cite{Fert2013}

These challenges might be overcome by realizing racetrack memory devices based on magnetic skyrmions.\cite{Fert2013, Tomasello2014, Zhang2015c, Muller2016}
They were observed for the first time in 2009\cite{Muhlbauer2009a} and theoretically discussed already more than 25 years ago.\cite{Pokrovsky1979, Bogdanov1989,Bogdanov1994}
Since 2009 skyrmions have been detected in various bulk materials\cite{Pfleiderer2010, Seki2012, Adams2012} and thin films.\cite{Yu2010, Yu2011b, Yu2012, Heinze2011, Onose2012, Romming2013}
They have also been shown to be stable up to room temperature, and skyrmions occur in different sizes.
Skyrmions are particularly interesting for device relevant systems due to their special properties:
i) Skyrmions are particle like and are usually repelled by smooth boundaries, so they do not touch the edges of the sample as domain walls always do;
ii) they are topologically non-trivial and therefore more stable than other magnetic textures;
iii) they can be efficiently manipulated by ultra-low electric currents,\cite{Jonietz2010, Schulz2012, Everschor2012a, Nagaosa2013, Litzius2017, Jiang2016} much smaller than the currents needed to move domain walls in magnetic wires; and
iv) the spacing between bits could be of the order of the skyrmion diameter, which allows for a much denser storage compared to domain walls.\cite{Fert2013}

To efficiently build high-performance skyrmion-based devices a reliable and controllable way to create skyrmions is needed.
So far several techniques to obtain single skyrmions have been proposed.\cite{Mohseni2013, Romming2013, Zhou2014, Li2014d, Jiang2015a, Zhou2015, Yuan2016, Muller2016a, Heinonen2016, Hrabec2016, Legrand2017, AxelHoffmannDPG2017, Yu2017a}
However, most of them either require specialized setups or artificially tuned parameters.
One recent example is the use of an inhomogeneous current distribution coupled to a chiral Dzyaloshinskii-Moriya interaction (DMI).\cite{Jiang2015a}
This set-up provides a rather uncontrolled skyrmion source where the skrymion creation is a ``random process'' similar to bubble creation in hydrodynamics.
Furthermore, most of the theoretical studies of the dynamics\cite{Han10, Iwasaki2013, Petrova2011a, Kalinkin2011, Liu2013, Iwasaki2014, Tchoe2012, Tretiakov2007, Iwasaki2014b, Rossler2006, Mochizuki2012, Koshibae2016, Heinonen2016, Leonov2016c} and creation \cite{Lin2013, Lin2014, Hrabec2016} of these spin textures imply or assume that the presence of a twisting  microscopic interactions, such as DMI, is required  \emph{for their creation}.

\begin{figure}[tb]
\centering
\includegraphics[width=0.5\columnwidth]{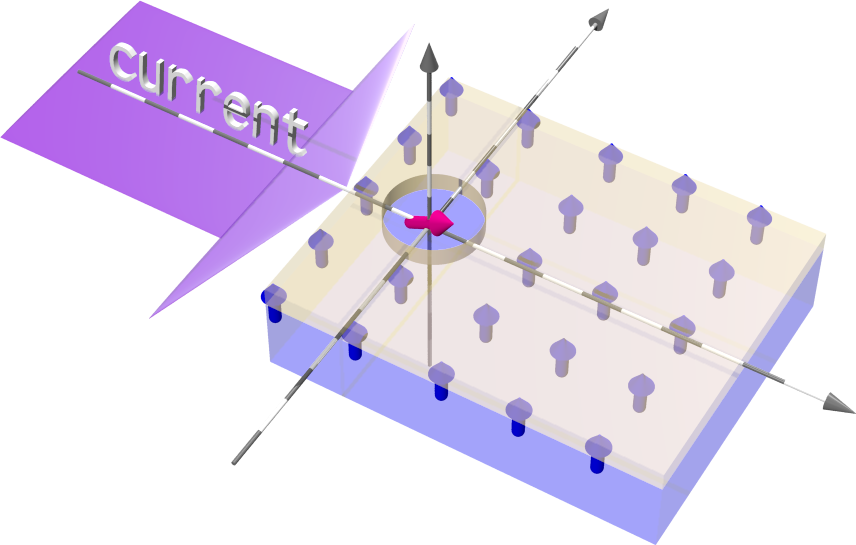}
\caption{%
Schematic experimental setup with perpendicular out-of-plane anisotropy in a bilayer thin-film structure, where the anisotropy is modified in a small area via locally modifying the top layer.
}
\label{fig:setup}
\end{figure}

\begin{figure}[tb]
\centering
\includegraphics[width=0.9\columnwidth]{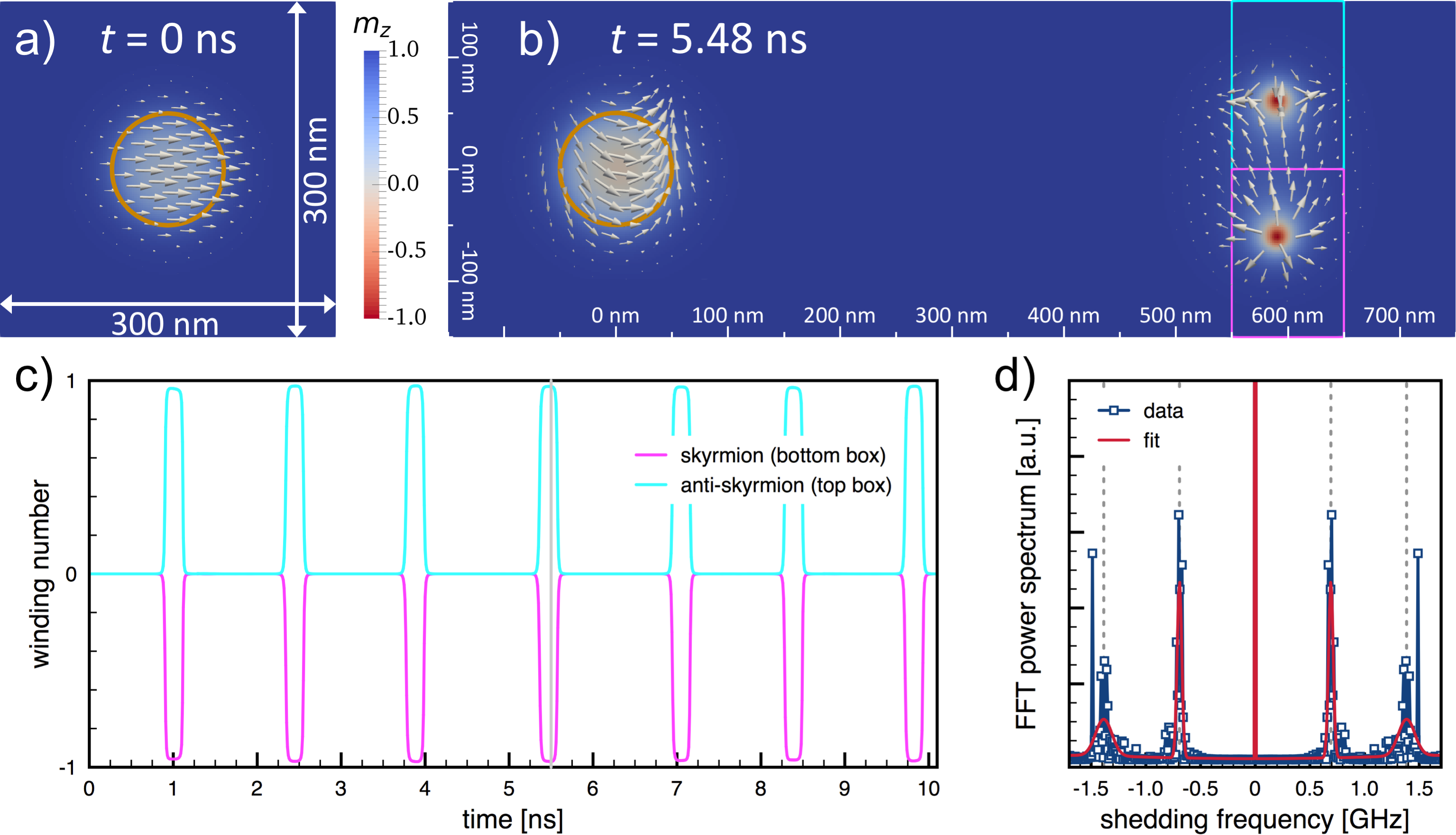}
\caption{%
Main results from micromagnetic simulations for the setup of Fig.~\ref{fig:setup} where we have chosen material parameters similar to CoCrPt thin films:\cite{Zheng2002, Roy2003, Navas2010, Kane2015}
$M_{s} = \SI{3e5}{\ampere\per\meter}$, $A_{\mathrm{ex}} = \SI{2e-11}{\joule\per\meter}$, $K_{u} = \SI{8e4}{\joule\per\cubic\meter}$, $K_{u}^{p} = 0$ and $\tilde{K}_{u}^{p} = \SI{1e3}{\joule\per\cubic\meter}$ \emph{inside} the pinning area (indicated by yellow circle) which has a radius of \SI{50}{\nano\meter}, and Gilbert damping of $\alpha = 0.05$, and $\beta = 0$. The applied current density is $\SI{5e12}{\ampere\per\square\meter}$.
\textbf{a)} Close-up view of the initial magnetization configuration close to the pinning center.
The color code indicates the $m_{z}$-component, while the arrows visualize the in-plane component of the magnetization.
\textbf{b)} Snapshot of the time evolution under an applied DC current at $t = \SI{5.48}{\nano\second}$.
The skyrmion/antiskyrmion pair exists over a long length of the order of a micrometer.
The cyan and magenta boxes indicates regions where we calculate the winding number of each topological object.
\textbf{c)} Winding number of the topological objects  as function of time computed over the regions indicated by the box regions in panel b).
\textbf{d)} Power spectrum of the Fourier transform of the winding number (blue data) and fit (red line) of the peaks up to second harmonic (indicated by dashed grey lines) with a shedding frequency of $f_{\mathrm{shed}} \approx \SI{0.68}{\giga\hertz}$.
}
\label{fig:continuous_mode}
\end{figure}

Here we present a mechanism to periodically produce magnetic textures in simple thin film geometries by means of a \emph{homogeneous} DC current and a spatially inhomogeneous magnetization without requiring any standard ``twisting'' interactions. To propose a concrete setup we consider an experimentally realizable pinning center, that creates the magnetic inhomogeneity, see Fig.~\ref{fig:setup}.
The magnetic textures can be created efficiently and controllably, as illustrated in Fig.~\ref{fig:continuous_mode}.
This mechanism relies physically on a local static loss of stability created by the combined interaction of current-induced spin-transfer torque and the pinning center leading to a bifurcation into a spatio-temporal period pattern.
It is similar to our recent work in one dimension leading to current induced domain wall production\cite{Sitte2016}, however in the two-dimensional case the situation is far more complex.
We demonstrate that it is possible to use this mechanism to create skyrmion/antiskyrmion pairs.
The inclusion of DMI interactions plays no major role in their creation process but does affect the subsequent dynamics of the magnetic textures and in particular ultimately stabilize one member of the pair with the preferred chirality.
A recent theoretical study focussing primarily on systems with DMI\cite{Stier2017} have also demonstrated the creation of skyrmion/antiskyrmion pairs, also verifying this in the absence of DMI.

This paper is structured as follows:
First we provide a simple physical picture of the skyrmion formation.
As an example, we explicitly describe a experimental setup in which we predict that skyrmions can be produced via a homogeneous DC current.
Then we discuss our main analytical and numerical results.
In particular, we consider skyrmion/antiskyrmion pair production and their time evolution and discuss the option of a pulse-operation mode of the device.
We first analyse the creation process in a setup without DMI interactions and later on include chiral interactions to study the different time evolution of the created textures.
At the end we discuss our results and allude to experiments that might have already observed the creation mechanism proposed in this paper.

\section{Physical picture of magnetic texture formation by DC currents}

As it follows from the analytics section described below, a DC current is able to produce magnetic textures once there is a magnetic inhomogeneity in the system.
To study a concrete and reproducible example, we focus on the setup illustrated in Fig.~\ref{fig:setup}, consisting of a metallic ferromagnetic film with a perpendicular uniaxial anisotropy where in a small region of the magnetic film the magnetization is tilted.
In this work we realize this tilting by a change in the magnetocrystalline anisotropy which acts as a pinning center.
We then apply a DC current in an in-plane direction.

The physical picture of the newly introduced mechanism for topological magnetic texture formation is as follows:
i) when ramping up the current strength the current  modifies the magnetization structure around the pinning center such that the area, where the magnetization is nonuniform, becomes elongated;
ii) the shape and size of the area of the nonuniform magnetization depends on the strength of the current and microscopic details of the sample.
However, since the current couples to the spatial gradient of the magnetization, it acts mainly on the magnetization in the nonuniform area, \textit{i.e.}, near the pinning center;
iii) increasing the current density further, above a critical current density $j_{c}$, the local static magnetic texture becomes unstable and the current-induced spin-transfer-torque pushes away the nucleated texture and effectively shed it from the pinning center;
iv) the vicinity of the pinning center is then somewhat restored to its initial state, and the process can restart, leading to a periodic shedding process if the current is kept constant above $j_{c}$.
Here we would like to stress that the current density above which the ferromagnetic ground state becomes unstable is higher than the current densities used to obtain the shedding.\cite{Bazaliy98, Shibata2005, Sitte2016}

The above statements i) can be deduced from the analytics part presented in Sec.~\ref{sec:analytics}, ii) are analogous to the one dimensional case\cite{Sitte2016, Shibata2005}, and iii) can be deduced from a more general perspective from hydrodynamic theory.\cite{Iacocca2016}
However what remains unclear from those general arguments is how these shedded magnetic textures look like.
To this end we have performed micromagnetic simulations and we find that different magnetic textures can be shedded (see Supplementary Material\cite{Supplement}).
In particular, for currents slightly above $j_c$ and a small enough pinning center it is possible to periodically shed skyrmion/antiskyrmion pairs.

In the process described above, no twisting interaction is needed.
The addition of DMI would not aid or hinder the periodic creation process of magnetic textures, but will of course be necessary in a different region of the device to insure long term stability of stored skyrmions, see Sec.~\ref{ssec:numerics_with_dmi}.
It is important to emphasize that the newly introduced periodic texture production mechanism is a generic and ubiquitous process: neither the directions of the anisotropies nor the details of how it is locally altered in strength or orientation at the pinning center are crucial.
However, in order to demonstrate the practical feasibility of the proposed generic mechanism, we have specified an experimental setup based on CoCrPt thin films\cite{Zheng2002, Roy2003, Navas2010, Kane2015} to observe the formation of skyrmions by uniform DC currents in common ferromagnetic materials, as shown in Fig.~\ref{fig:setup}.
Locally modifying the top layer in a small region might reduce the out-of-plane anisotropy leading to an in-plane tilting of the magnetization within this region, as a practical implementation of the required local pinning center.

\section{Model for magnetization dynamics}
\label{sec:model}

To describe the current-induced magnetization dynamics of the considered ferromagnetic thin film we use the Landau-Lifshitz-Gilbert equation for the unit vector field $\hat{\vect{M}}$ (Ref.~\citenum{Zhang2004}) generalized to include spin-torque effects due to the electric current:
\begin{equation}
(\partial_{t} + v_{s} \partial_{x}) \hat{\vect{M}} = -\gamma \hat{\vect{M}} \times \vect{B}_{\mathrm{eff}} + \alpha \hat{\vect{M}} \times \biggl( \partial_{t} + \frac{\beta}{\alpha} v_{s} \partial_{x} \biggr) \hat{\vect{M}},
\end{equation}
where $\gamma$ is the gyromagnetic ratio, and $\alpha$ and $\beta$ are {the} dimensionless Gilbert {damping} and non-adiabatic {spin-transfer-torque} parameters.
The effective magnetic field is given by $\vect{B}_{\mathrm{eff}} = -M_{s}^{-1} (\delta F[\hat{\vect{M}}]/\delta \hat{\vect{M}})$, where $F[\hat{\vect{M}}]$ describes the free energy of the system and $M_{s}$ is the saturation magnetization.
We decompose the free energy $F$ into two parts $F = F_{0} + F_{\mathrm{twist}}$ where $F_{0}$ consists of isotropic exchange, anisotropy term and dipolar interactions:
\begin{equation}
\label{eq:energy}
F_{0}[\hat{\vect{M}}] = \int \biggl[ A_{\mathrm{ex}} (\nabla \hat{\vect{M}})^{2} + \Pi(\hat{\vect{M}}) - \frac{\mu_0}{2} M_{s} \hat{\vect{M}} \cdot \vect{H}_d (\hat{\vect{M}}) \biggr] dV,
\end{equation}
where $A_{\mathrm{ex}}$ is the exchange constant, $\Pi(\hat{\vect{M}})$ describes the functional form of the anisotropy energy and the last term describes the dipolar interactions.
$F_{\mathrm{twist}}$ describes a twisting interaction like the DMI interaction, which we set to zero in the first part of the numerics.
Thereby we show that such a twisting term is not crucial for the creation of the magnetic textures.
In the latter part of the numerics we do, however, explicitly consider different twisting terms.
The applied uniform DC current along the $x$ direction enters the equation via the effective spin velocity, $v_{s} = \xi j$ with $\xi = g P \mu_{B}/(2 e M_{s})$, where $g$ is the g-factor, $P$ is the current polarization, $\mu_{B}$ is the Bohr magneton, $e$ is the electron charge, and $j$ is the current.

\section{Results: Analytics}
\label{sec:analytics}

In this section we demonstrate the general aspects of this new type of magnetic texture production in thin films.
We show below that spin textures are periodically created above a critical current $j_{c}$, with a period $T \sim (j - j_{c})^{-1/2}$.
This periodic texture production is quite general and does not depend on the details of the microscopic Hamiltonian for a large class of magnetic systems.
However, the details of the process, such as the value of the critical current $j_{c}$ or the prefactor of the periodic scaling, do depend on the microscopic details of the Hamiltonian.
Recently we have analysed the same mechanism for one-dimensional nano-wires,\cite{Sitte2016} where a pinning center leads to current-induced periodic domain wall production.
The one-dimensional problem is analytically solvable including the calculation of the shape of the emerging magnetic texture of the domain wall, whereas the situation in two dimensions is far more complex.
Analytically it is still possible to derive the shedding period based on generalized arguments as shown below.
By general topology arguments one can infer that the winding number during the creation process is conserved, but the precise shape of the produced magnetic textures cannot be calculated analytically.

The only assumptions that enter our analytical calculations are that
i) the magnetic free energy density of the system is translationally invariant outside the pinning area, and
ii) that we neglect the non-adiabatic spin-transfer torque term.
The critical scaling related to the type of instability may be transformed by non-adiabatic corrections, but the general finding of periodic production of spin textures by DC currents is expected to remain valid.
Below we discuss the main analytical strategy and findings, whereas a full derivation of our analytical results can be found in the Supplementary Material.\cite{Supplement}

\begin{figure}
\includegraphics[width=.7\columnwidth]{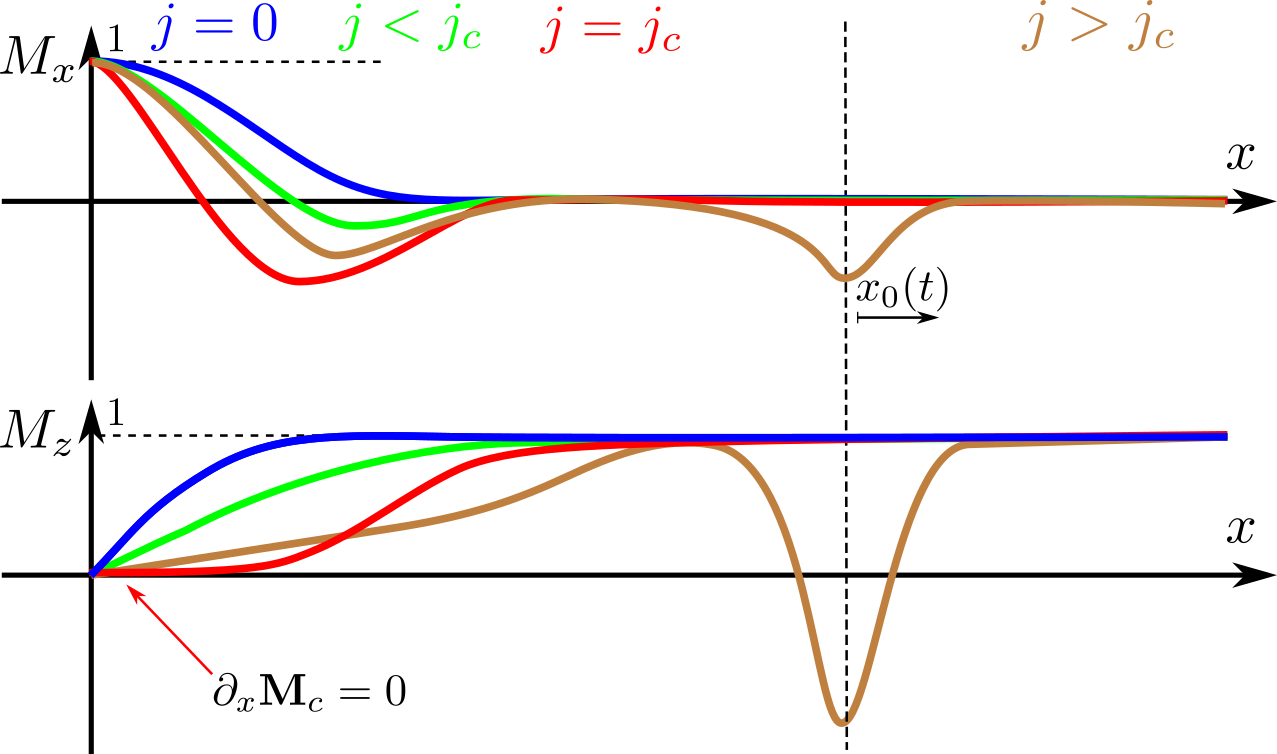}
\caption{%
Schematic plot for which the local pinning direction is along the $x$ direction, and the global pinning direction is along the $z$ direction, \textit{i.e.}, $M_{x}(x = 0) = 1$ for all current strengths.
At the critical current all derivatives at $x=0$ are zero.
}
\label{fig:schematic_plot}
\end{figure}

The critical current density is defined by the instance when the magnetic texture is about to rip off.
Because the current density couples to the gradient of the magnetization via the spin-transfer torque, the critical current density is defined by the gradient of the magnetization profile at the pinning center going to zero:
\begin{equation}
\label{eq:jc_main}
\partial_{x} \hat{\vect{M}}_{c}(j_{c}; \vect{r}=0) = 0,
\end{equation}
where $\hat{\vect{M}}_{c}(\vect{r})$ is the magnetization profile obtained by solving the Landau-Lifshitz-Gilbert equation for a current density $j_{c}$.
A sketch of the magnetization profile for different current strengths is shown in Fig.~\ref{fig:schematic_plot}.

The period at which magnetic textures are created is derived by combining two crucial arguments: one arises from the ``just still static limit'' ($j_{0} \lesssim j_{c}$) and the other one from the ``just dynamic limit'' ($j \gtrsim j_{c}$).
For the first one, the crucial point to note is that at a current $j_{0}$ slightly smaller than the critical current ($j_{0} \lesssim j_{c}$) the magnetization profile will not differ too much from the critical one besides being shifted by a small distance $x_{0}$: $\hat{\vect{M}}_{0}(\vect{r}) \equiv \hat{\vect{M}}_{c}(\vect{r} - x_{0} \vect{e}_{x}) + \delta \hat{\vect{M}}$.
Solving the LLG in the static limit with this ansatz yields
\begin{equation}
\label{eq:j_below_jc_main}
j_{c} - j_{0} \sim x_{0}^{2}.
\end{equation}
This equation establishes the relation between the spatial shift of the critical solution, $x_{0}$, and the current $j_{0}$.
As a check we note that i) for $x_{0} = 0$, $j_{0}$ and the critical current $j_{c}$ coincide, and that ii) the difference $j_{c} - j_{0}$ is second order in $x_{0}$, as expected from perturbation theory arguments.

For currents just larger than the critical current $j_{c}$, the static solutions are unstable.
Therefore, to obtain information about the dynamics of the magnetization configuration, a justified ansatz for the magnetization at an applied current $j$ is a sum of $\vect{M}_{0}$ (with a time dependent shift $x_{0}$) and a small perturbation: $\vect{M}(\vect{r}, t) =   \vect{M}_{0}(\vect{r}; x_{0}(t)) + \vect{m}(\vect{r}, t)$.
Here, $\vect{M}_{0}(\vect{r})$ is the static magnetization configuration for the shift $x_{0}$ which can be parametrized by a current $j_{0}$ using Eq.~\eqref{eq:j_below_jc_main}.
For this ansatz, the Landau-Lifshitz-Gilbert equation has a direct solution with $\vect{m} \equiv 0$ and
\begin{equation}
\label{eq:j_above_jc_main}
\partial_{t} x_{0} = j - j_{0},
\end{equation}
implying that the velocity of the magnetic texture is proportional to the applied current.

Solving Eq.~\eqref{eq:j_below_jc_main} for $j_{0}$ and inserting it in Eq.~\eqref{eq:j_above_jc_main} yields the period of the magnetic texture formation:
\begin{equation}
\label{eq:period_main}
T \sim (j - j_{c})^{-1/2}.
\end{equation}

\section{Results: Numerics}

We have performed micromagnetic simulations with various setups and various configurations, using both MicroMagnum\cite{MicroMagnum} including additional self-written software extensions as well as MuMax3.\cite{Vansteenkiste2014}
For details and parameters used see the Methods Section (Sec.~\ref{sec:MicromagneticSim}) and for a summary table of the parameters see the Supplementary Material.\cite{Supplement}
In the first part we focus on the periodic texture formation, in particular on the skyrmion and anti-skyrmion creation itself and try to isolate the mechanism to show that DMI is not important for their creation.
We analyse the shedding and the time evolution.
After that we consider the shedding process in the presence of the generalized anisotropic DMI\cite{Camosi2017, Hoffmann2017} which after the skyrmion and antiskyrmion pairs are created, support the stabilization of the skyrmion or the antiskyrmion, depending on the details and the signs of the entries of the DMI tensor, respectively.

In the following we will present our simulation results for a DC current which is ramped up to a constant value above $j_{c}$, where we have used the following form of the anisotropy term:
\begin{equation}
\Pi(\hat{\vect{M}}) =
  \begin{cases}
    K_{u} (1 - M_{z}^{2}), & \text{for the film},\\
    K_{u}^{p} (1 - M_{z}^{2}) + \tilde{K}_{u}^{p} (1 - M_{x}^{2}), & \text{in the pinning center}.
  \end{cases}
\end{equation}
We assume that in the pinning center the anisotropy strength $K_{u}^{p}$ is reduced, \textit{i.e.}, $K_{u}^{p} < K_{u}$, such that the corresponding effective anisotropy in the ultrathin film limit induced by the dipolar interactions ($K_{\mathrm{eff}} = K - (\mu_{0}/2) M_{s}^2$) is positive (negative) for $K_{u}$ ($K_{u}^{p}$).
This effectively leads to a tilting of the magnetic texture into the plane within the pinning region.
Since in a real material there is always a small uniaxial anisotropy also in an in-plane direction, we have taken a small rotational symmetry breaking term into account, tilting the magnetic texture along $x$ direction.
This corresponds for example to the setup depicted schematically in Fig.~\ref{fig:setup}.

In the Supplementary Material we present additional results:
i) for simulations even without dipolar interactions, in which a skyrmion and an antiskyrmion are energetically fully equivalent wherefore the creation and decay process for skyrmions and antiskyrmions is fully symmetric;
ii) for the creation of different magnetic textures
and iii) a pulse operation mode of the "device".

\subsection{Numerics without DMI}
A typical relaxed magnetization configuration is shown in Fig.~\ref{fig:continuous_mode}~(a).
In the continous operation mode, a DC current above the critical one can generate a periodically spin textures.
For a large set of parameters these spin textures evolve into skyrmion/antiskyrmion pairs.
A typical snapshot for the time evolution of the magnetization under a (continuous) DC current is presented in Fig.~\ref{fig:continuous_mode}~(b).
To establish the topological nature of the skyrmion and antiskyrmion we have calculated as a function of time the winding numbers over spatial regions indicated by the boxes in cyan and magenta.
The results presented in panel (c) clearly show a period creation of skyrmion/ antiskyrmion pairs with winding numbers $\pm 1$.
The shedding frequency for the simulated material structure with applied current density of \SI{5e12}{\ampere\per\square\meter} can be extracted from the power spectrum of the Fourier transformed data, $\omega_{\mathrm{shed}} \approx \SI{0.68}{\giga\hertz}$ corresponding to a shedding period of approximately\ $T_{\mathrm{shed}}\approx \SI{1.47}{\nano\second}$ [cf.\ panel (d)].
For this setup, we have used $M_{s} = \SI{3e5}{\ampere\per\meter}$, $A_{\mathrm{ex}} = \SI{2e-11}{\joule\per\meter}$, $K_{u} = \SI{8e4}{\joule\per\cubic\meter}$, $K_{u}^{p} = 0$ and $\tilde{K}_{u}^{p} = \SI{1e3}{\joule\per\cubic\meter}$ \emph{inside} the pinning area which has a radius of \SI{50}{\nano\meter}, and Gilbert damping of $\alpha = 0.05$, and $\beta = 0$.
More information on the numerical part can be found in the Methods Section, Sec.~\ref{sec:MicromagneticSim}.

\subsubsection{Shedding of skyrmion/antiskyrmion pairs}

\begin{figure}[tb]
\centering
\includegraphics[width=.9\columnwidth]{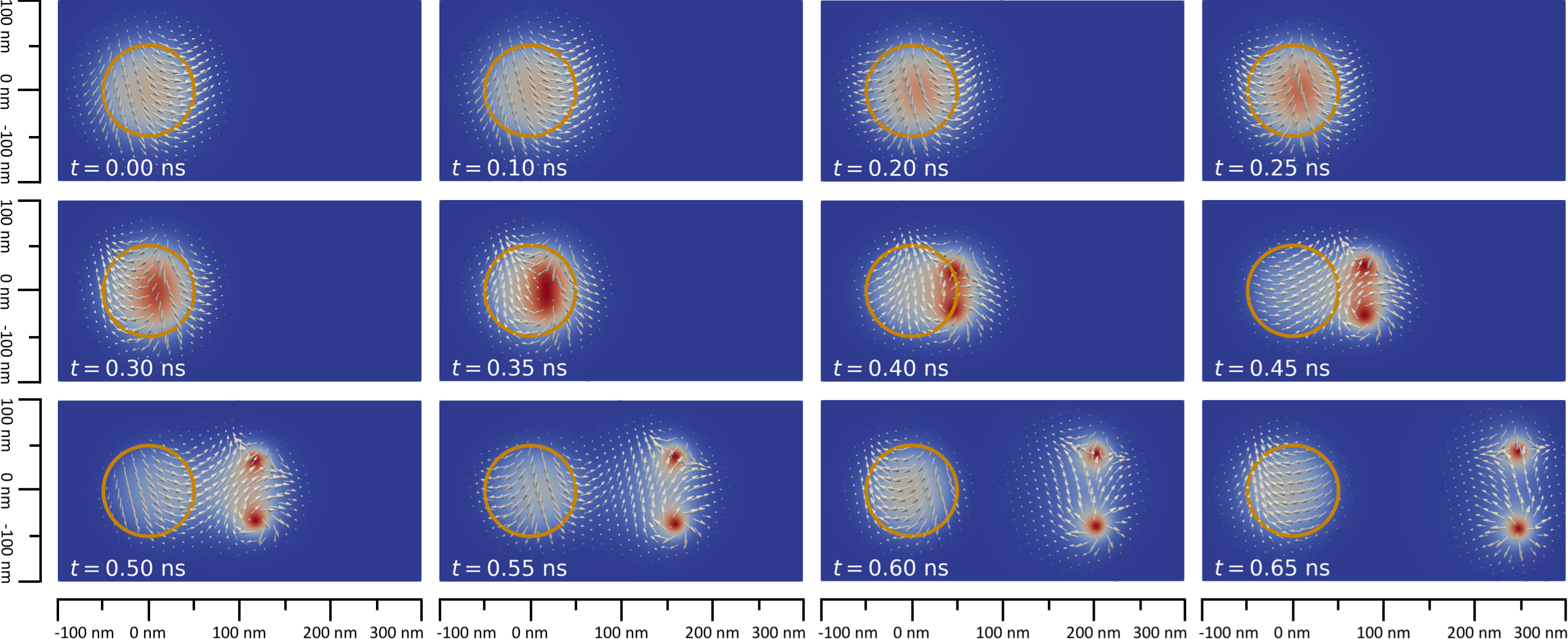}
\caption{%
Details of the shedding process of the time evolution of a skyrmion/antiskyrmion pair shown in Fig.~\ref{fig:continuous_mode}.
Shown is the $z$ component as a color code, and the arrows visualize the in-plane components.
For this setup, we use the same parameters as in Fig.~\ref{fig:continuous_mode}: $M_{s} = \SI{3e5}{\ampere\per\meter}$, $A_{\mathrm{ex}} = \SI{2e-11}{\joule\per\meter}$, $K_{u} = \SI{8e4}{\joule\per\cubic\meter}$, $K_{u}^{p} = 0$ and $\tilde{K}_{u}^{p} = \SI{1e3}{\joule\per\cubic\meter}$ \emph{inside} the pinning area which has a radius of \SI{50}{\nano\meter}, Gilbert damping of $\alpha = 0.05$, and $\beta = 0$.
The applied current density is $\SI{5e12}{\ampere\per\square\meter}$.
}
\label{fig:shedding}
\end{figure}

For currents less than the critical current, the magnetic texture around the pinning center elongates until above the critical current it breaks off and a first skyrmion/antiskyrmion pair is formed, which travels along the magnetic film.
The details of the shedding process are shown in Fig.~\ref{fig:shedding}, where we show the profiles of the pinning center up to $\SI{0.65}{\nano\second}$.
The observation that the created topological texture comes in a pair with opposite topological charge reflects the fact that during the creation process the topological charge is conserved, with initial configuration having no topological charge.

\subsubsection{Time-evolution of a skyrmion/antiskyrmion pair}

\begin{figure}[tb]
\centering
\includegraphics[width=.95\columnwidth]{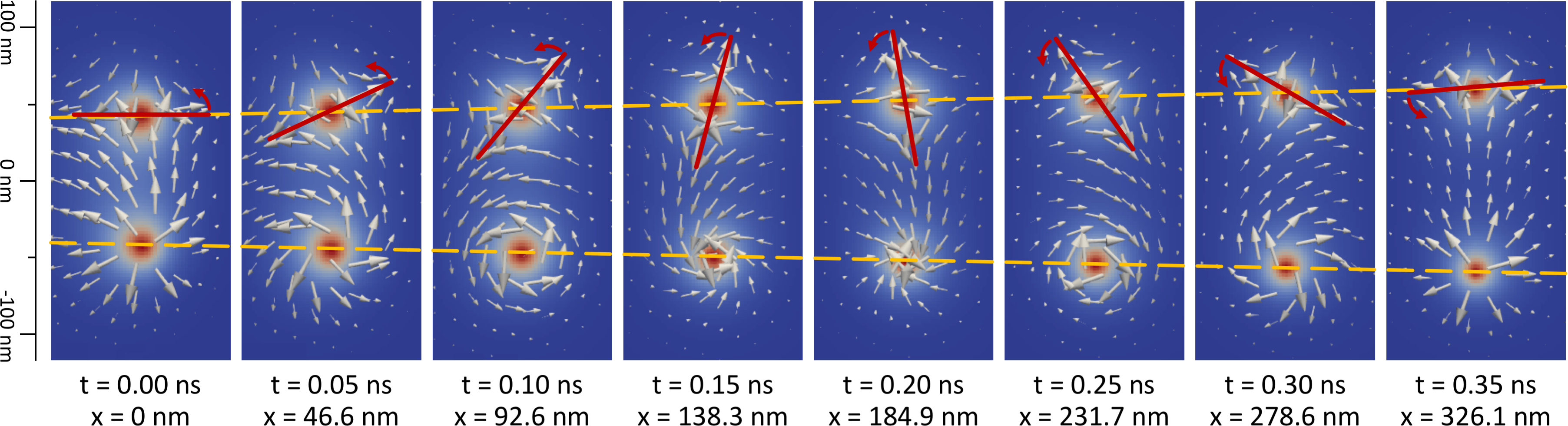}
\caption{%
Detailed time evolution of a skyrmion/antiskyrmion pair in time steps of \SI{0.05}{\nano\second}.
The skyrmion (lower configuration) oscillates continuously in time from a Bloch to a N{\'e}el skyrmion and from positive to negative chirality.\cite{Zhou2015}
Here all spins globally rotate counterclockwise in time.
The same applies for the antiskyrmion (upper configuration), leading to a continous counterclockwise rotation of the red line (serving as guide to the eye) at which the spins are pointing away from the center of the antiskyrmion.
The oscillation frequency is $f_{\mathrm{osc}} \approx 1/\SI{0.34}{\nano\second} \approx \SI{2.9}{\giga\hertz}$.
The dashed yellow lines visualize the drift in opposite directions in $y$ direction due to opposite Magnus forces for skyrmion and antiskyrmion.
For this setup, we used the same parameters as in Fig.~\ref{fig:continuous_mode}: $M_{s} = \SI{3e5}{\ampere\per\meter}$, $A_{\mathrm{ex}} = \SI{2e-11}{\joule\per\meter}$, $K_{u} = \SI{8e4}{\joule\per\cubic\meter}$, $K_{u}^{p} = 0$ and $\tilde{K}_{u}^{p} = \SI{1e3}{\joule\per\cubic\meter}$ \emph{inside} the pinning area which has a radius of \SI{50}{\nano\meter}, and Gilbert damping of $\alpha = 0.05$, and $\beta = 0$.
Applied current is $\SI{5e12}{\ampere\per\square\meter}$.
}
\label{fig:pair_evolution}
\end{figure}

The skyrmion/antiskyrmion pair begins to move away from the pinning center and we observe in our simulations that the distance between the anti-skyrmion and skyrmion increases at a rate proportional to $\alpha$ as expected (with $\beta = 0$ in our simulations).
The evolution of the pair has an oscillatory character related to the fact that, in equilibrium, the individual structures are not stable.
This is shown in detail in Fig.~\ref{fig:pair_evolution}.
The dynamical stabilization due to the current leads to the global continuous precession of all spins around the $z$ axis.
This results in the skyrmion (lower magnetic texture) oscillating continuously between N{\'e}el and Bloch skyrmion type and between negative and positive chirality, as discussed in Ref.~\citenum{Zhou2015}, and to the anti-skyrmion (upper magnetic texture) rotating its orientation counterclockwise.
For the modelled setup we obtain an oscillation frequency of $f_{\mathrm{osc}} \approx 1/\SI{0.34}{\nano\second} \approx \SI{2.9}{\giga\hertz}$ being about four times faster than the shedding frequency.
This oscillatory behaviour does depend on the details of the specific sample and therefore we do not explore it further, since it does not affect the production or control of the texture in the considered time and length scale.

\subsubsection{Decay of magnetic textures}

In presence of a non-zero Gilbert damping, the created topological textures which are dynamical solitons \cite{kosevich1990, Zhou2015}, progressively decay (their radius shrinks) at a rate proportional to $\alpha$.
They will ultimately collapse in a finite time due to imperfect topological protection on a lattice, unless they reach an area with non zero chiral interaction at which place either the skyrmion or the antiskyrmion is meta-stable allowing for longer term storage.

\begin{figure}[tb]
\centering
\includegraphics[width=0.9\textwidth]{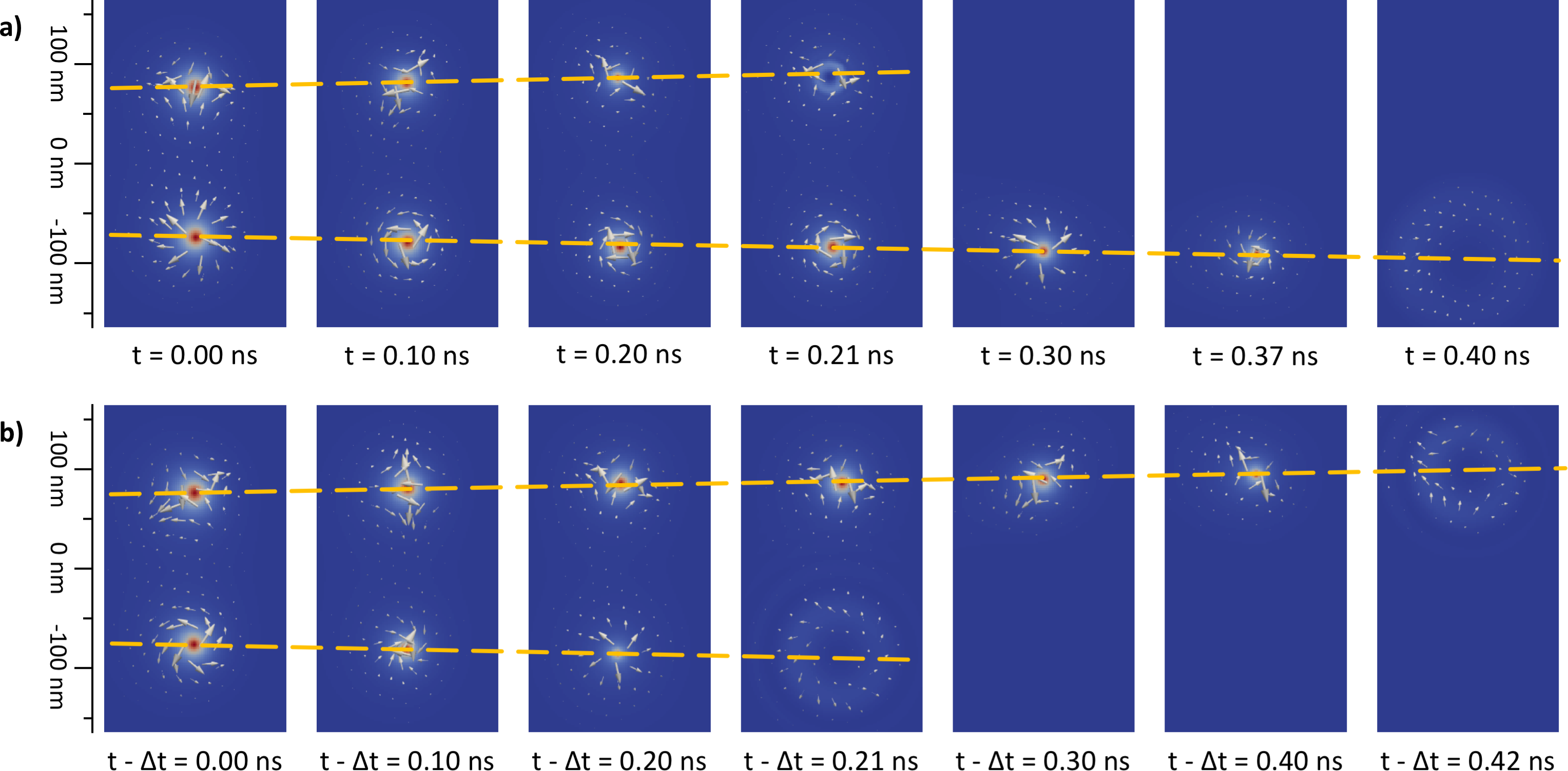}
\caption{%
Snapshot of the time evolution of a skyrmion/antiskyrmion pair where in the upper panel a) (lower panel b)) the antiskyrmion (skyrmion) decays faster than the skyrmion (antiskyrmion) due to interactions with spin waves generated by previous skyrmion/antiskyrmion pairs.
Note that the top and bottom panel are taken from the same simulation, but from different pairs created at different times.
The event shown in panel b) occurs about $\Delta t= \SI{3}{\nano\second}$ later.
Furthermore, the two events are not fully symmetric and do depend on the history of the system.
For this setup, we used the same parameters as in Fig.~\ref{fig:continuous_mode}: $M_{s} = \SI{3e5}{\ampere\per\meter}$, $A_{\mathrm{ex}} = \SI{2e-11}{\joule\per\meter}$, $K_{u} = \SI{8e4}{\joule\per\cubic\meter}$, $K_{u}^{p} = 0$ and $\tilde{K}_{u}^{p} = \SI{1e3}{\joule\per\cubic\meter}$ \emph{inside} the pinning area which has a radius of \SI{50}{\nano\meter}, and Gilbert damping of $\alpha = 0.05$, and $\beta = 0$.
Applied current is $\SI{5e12}{\ampere\per\square\meter}$.
}
\label{fig:decay}
\end{figure}

In the presented examples the skyrmion and anti-skyrmion do not decay at the same time.
In the presence of dipolar interactions, which do not break the rotational symmetry, a skyrmion is energetically favored compared to the antiskyrmion, resulting in principle in a slightly larger life time.
However, the shedding process itself, the oscillation of the magnetic textures, as well as their decay create a non-negligible amount of spin waves, which in turn influence the time evolution of the magnetic textures in the system.
Therefore, a complete understanding of a certain decay process does depend on the history of the system.
In our micromagnetic simulations we indeed observe instances where either the skyrmion or the antiskyrmion decays first as can be seen in Fig.~\ref{fig:decay}, where the top and bottom row are taken from the same simulation just at different times.
In the Supplementary Material\cite{Supplement} we consider also a system without dipolar interactions where neither skyrmion nor antiskyrmion are favoured and we observe that they decay at the same time.

\subsection{Numerics with anisotropic DMI}
\label{ssec:numerics_with_dmi}

The simulations presented above were intentionally performed without chiral interactions to explore the creation mechanism.
For the stabilization of chiral magnetic textures, however, an inversion asymmetric interaction is needed.
In the following, we consider the creation mechanism in the presence of anisotropic N\'eel DMI:\cite{Camosi2017, Hoffmann2017}
\begin{equation}
F_{\mathrm{twist}} = \int \biggl[ D_{1} \biggl( M_{x} \frac{\partial M_{z}}{\partial x} - M_{z} \frac{\partial M_{x}}{\partial x} \biggr) + D_{2} \biggl( M_{y} \frac{\partial M_{z}}{\partial y} - M_{z} \frac{\partial M_{y}}{\partial y} \biggr) \biggr] dV.
\end{equation}
For $D_{1} = D_{2} = D$, the twisting energy $F_{\mathrm{twist}}$ describes the standard N\'eel DMI\cite{Thiaville2012} favouring magnetic skyrmions.
For $\sgn(D_{1}/D_{2}) = +1$ a skyrmion like configuration is still energetically more favourable, whereas for $\sgn(D_{1}/D_{2}) = -1$ an antiskyrmion like configuration will have a lower energy compared to a skyrmion like texture.\cite{Camosi2017, Hoffmann2017}
Anti-skyrmions have been studied, but until recently they have often been considered as unstable objects.\cite{Koshibae2016}
More recent works show that anti-skyrmions can also occur as (meta-)stable states.\cite{Dupe2016a, Camosi2017, Hoffmann2017}

\begin{figure}[tb]
\centering
\includegraphics[width=0.9\columnwidth]{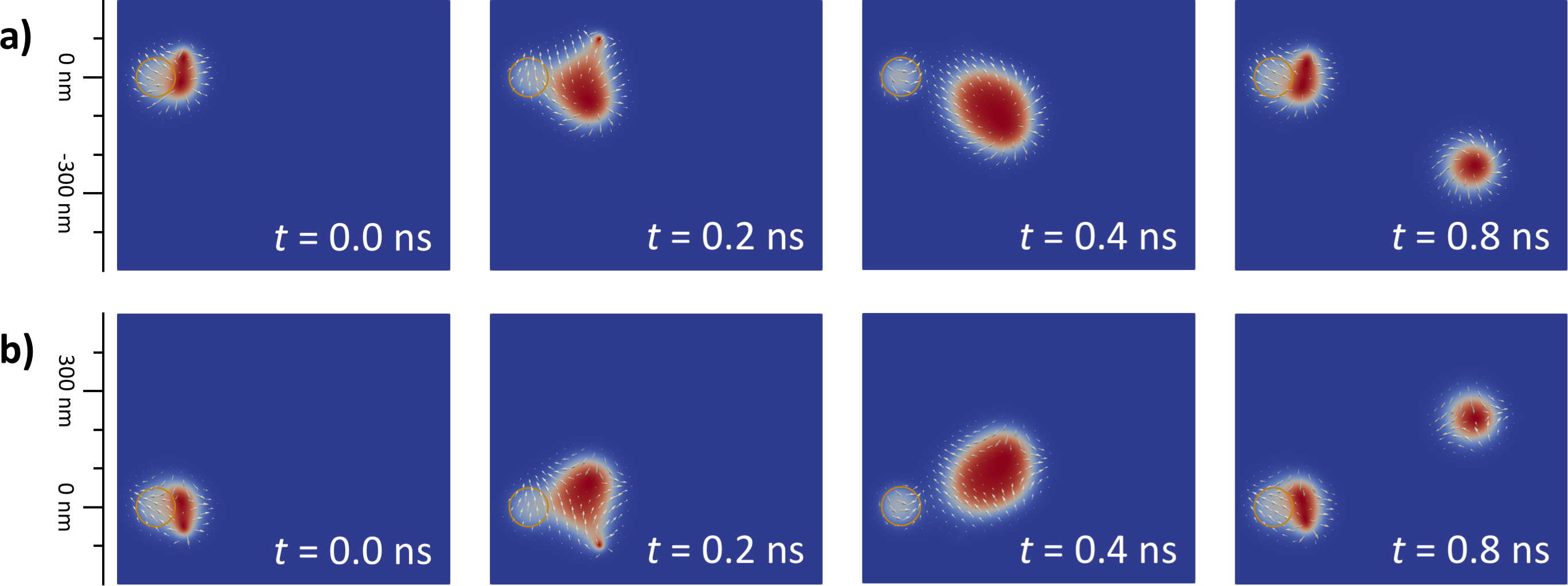}
\caption{
Upper row: Time evolution for $D_{1} = D_{2} = \SI{5e-4}{\joule\per\square\meter}$ corresponding to the DC current density $j = \SI{3.25e12}{\ampere\per\square\meter}$.
Again there is a production of skyrmion/antiskyrmion pairs, however during time evolution the skyrmion texture is preferred.
Lower row: Time evolution for $D_{1} = -D_{2} = \SI{5e-4}{\joule\per\square\meter}$.
Here, in contrast to normal DMI, the anti-skyrmion texture is preferred by the anti-DMI.
}
\label{fig:continuous_modeDMI}
\end{figure}

Simulation results with anisotropic DMI are shown in Fig.~\ref{fig:continuous_modeDMI}.
Here, skyrmion and anti-skyrmion pairs are created as previously, however, the time evolution for the pairs is now influenced by the DMI.
In the upper (lower) row in Fig.~\ref{fig:continuous_modeDMI} we considered a DMI with $D_{1} = \SI{5e-4}{\joule\per\square\meter}$ and $D_{2} = \pm D_{1}$, which optimally favour a symmetric skyrmion (antiskyrmion).

\section{Discussion}

The process that we have described is an alternative way to controllably produce skyrmions and also antiskyrmions periodically.
Whereas the process itself is quite general given a set of simple restrictions (a pinned magnetic inhomogeneity in the system caused by, for example, a local change in the anisotropy), the specific values of the critical current are set by the material properties.
For the examples chosen the current densities are typical of the spin-transfer torque process.
Although this implies the need for large current densities, leading to possible Joule heating, this heating can be mitigated experimentally by applying pulses, see Supplementary Material.\cite{Supplement}
Such heating does not alter the general conclusions and just renormalizes some of the material-dependent parameters.

The incorporation of a twisting interaction like (anisotropic) DMI is not essential for the physics of the skyrmion/antiskyrmion pair \emph{creation} process described here.
It affects the value of the critical current and the shape of textures generated.
In addition it helps towards a stabilization for the created (anti-)skyrmion texture in the static case.
Hence, there is no contradiction with the Derrick-Hobart theorem\cite{Hobart1963, Derrick1964} that states that there is no nontrivial localized static solution for skyrmion without chiral interactions.
In our process, which we believe is similar to the one obtained in Ref.~\citenum{Stier2017}, dynamic skyrmions and antiskyrmions are being produced, and could be stabilized at a later stage in regions with non zero (anisotropic) DMI.

In prior studies, several techniques to obtain single skyrmions have been proposed.\cite{Mohseni2013, Romming2013, Zhou2014, Li2014d, Jiang2015a, Zhou2015, Yuan2016, Muller2016a}
However, most of them either require specialized setups or artificially tuned parameters.
One recent example is the use of an inhomogeneous current distribution in the presence of DMI.\cite{Jiang2015a}
It is also possible to create magnetic skyrmions by an electric current in a sample with a suitable kink subject to DM interactions due to the divergence of the magnetization in the corner.\cite{Iwasaki2013}
In particular, here the authors observe an asymmetry in the current, meaning that they do not create skyrmions when the current is flowing in the opposite direction.
Recent numerical simulation studies show that a skyrmion can be created by the local injection of a spin-polarized current perpendicular to the plane.\cite{Sampaio2013,Zhou2015}
In Ref.~\citenum{Zhou2015} the authors also produce skyrmions in a setup without DMI, however they require a specialized setup with a nanocontact and the process is not connected to the periodic shedding created by the interaction between the current and the locally modified anisotropy as we propose here.

While these processes will be explored in the near future by experimental groups in search for efficient ways of controllably creating skyrmions, we propose a more general method that is applicable to systems readily available to a wider group of experimentalist.

We believe that the creation process discussed here has already been observed in recent experiment in variant setups, e.g. an injection of skyrmions is by DC current pulses.\cite{Hrabec2016, Legrand2017, AxelHoffmannDPG2017}
In Ref.~\citenum{Hrabec2016} the authors report a nucleation of skyrmions with a current pulse at the tip of a needle-like current injecting electrode.
Here, the Oersted field near the tip may act as a dynamically defined pinning center similar to what we consider here.
In Ref.~\citenum{Legrand2017}, we speculate that the grains in the sample are responsible for creating local inhomogeneities in the magnetization, i.e. induce pinning centers. When a dc-current is applied, at those points skyrmion and anti-skyrmion pairs will be created.
Since in all of these works\cite{Hrabec2016, Legrand2017, AxelHoffmannDPG2017} DMI is present in the considered systems, the antiskyrmions will die quickly on timescales that are not in the experimental accessible range in the measurement setup that have been reported so far.

To experimentally verify our prediction we suggest the following:
i) One of the key aspects of our mechanism is that it generates a periodic shedding of magnetic textures with a period that can be tuned by current strength, when applying a dc-current. In the above mentioned experiments, current pulses have been applied. When increasing the length of the current pulses towards having effectively a dc-current, it might be possible to measure the current dependent period.
ii) The other key aspect in our theory is the magnetic inhomogeneity. By increase locally the inhomogeneities at a certain point of the sample we predict that skyrmions can be produced in this region already with lower current densities. Examples to do so would be to have a small area where the sample is thinner or thicker. Alternatively, one can put a magnetic impurity into the system but the effect of the magnetic impurity may be harder to control.

\section{Conclusions}

In this work we have demonstrated that it is possible to create magnetic textures by uniform DC electric currents without the need of a magnetic field or any standard ``twisting'' interactions like Dzyaloshinskii-Moriya.
We have shown that skyrmion and antiskyrmion pairs can be created efficiently, controllably and periodically where the period can be tuned by the applied current strength.
We analytically derived, that in the limit of vanishing non-adiabatic torque, the period $T$ of production near the critical current has a dependence $T \sim (j - j_{c})^{-1/2}$, arising from quite general grounds based on simple symmetry consideration of the governing dynamic equations.
In principle, within a pulse operation mode skyrmions can be produced on demand.
This provides a new avenue to study the \emph{creation} of topological magnetic textures by electric means in simple geometries.
Adding DMI helps to stabilize magnetic skyrmions.

\section{Micromagnetic simulations}
\label{sec:MicromagneticSim}

Micromagnetic simulations were perfomed based on MicroMagnum\cite{MicroMagnum} including additional self-written software extensions and MuMax3.\cite{Vansteenkiste2014}
In the simulations shown in this article we have simulated a quasi-two-dimensional thin film with lateral dimensions $\SI{2}{\micro\meter} \times \SI{1}{\micro\meter} \times \SI{1}{\nano\meter}$ consisting of $1000 \times 500 \times 1$ cells in $x$, $y$, and $z$ direction, respectively, which corresponds to discretization length of $\SI{2}{\nano\meter}$ in $x$ and $y$ directions and $\SI{1}{\nano\meter}$ in $z$ direction.
For the plots shown in this manuscript we have chosen material parameters similar to CoCrPt thin films with perpendicular magnetic anisotropy:\cite{Zheng2002, Roy2003, Navas2010, Kane2015}
$M_{s} = \SI{3e5}{\ampere\per\meter}$, $A_{\mathrm{ex}} = \SI{2e-11}{\joule\per\meter}$, and out-of-plane anisotropy $K_{u} = \SI{2e4}{\joule\per\cubic\meter}$ with anisotropy energy functional $\Pi(\vect{M}) = 1 - M_{z}^{2}$.
The pinning center is modeled as a cylinder along $z$ direction with a radius of $\SI{50}{\nano\meter}$ and reduced anisotropy $K_{u} = 0$ in $z$ direction, but $\tilde{K}_{u}^{p} = \SI{1e3}{\joule\per\cubic\meter}$ in $x$ direction breaking the in-plane rotational symmetry.

Before applying any DC currents we have first relaxed the system to the ground state using a large Gilbert damping parameter of $\alpha = 0.25$.
Subsequent simulations were performed with $\alpha = 0.05$, except when noted otherwise.
In all simulations we have set the non-adiabatic spin-torque parameter to zero, $\beta = 0$.
We have checked that the results are independent of the grid sizes.

\section*{Acknowledgments}

We are grateful to M.~Thorwart, M.~Stier, O.~Gomonay, K.~Litzius, A.~Rosch, and K.~Hals for discussions.
Ar.~A.\ is also very grateful for the warm hospitality of the supporting staff of the INSPIRE group at Johannes Gutenberg-Universit{\"a}t, Mainz, Germany.
We acknowledge the funding from the Alexander von Humboldt Foundation,  the Transregional Collaborative Research Center (SFB/TRR) 173 Spin+X, and the ERC Synergy Grant SC2 (No.\ 610115).
K.~E.-S.\ acknowledges funding from the German Research Foundation (DFG) under the Project No.\ EV 196/2-1.

\section*{Author Contributions}
Ar.~A., K.~E.-S., T.~V., and J.~S.\ did the planning of the project.
Ar.~A.\ initiated the project and performed the analytic calculations.
T.~V.\ proposed device configurations.
M.~S., and K.E.-S.\ performed the numerical calculations.
All the authors contributed to the writing of the manuscript.

\bibliography{2017-06-10_Mendeley,references}

\section*{Additional Information}

Supplementary Information accompanies this paper.

\clearpage

\setcounter{equation}{0}
\setcounter{figure}{0}
\setcounter{section}{0}
\setcounter{table}{0}

\renewcommand{\theequation}{S\arabic{equation}}%
\renewcommand{\thefigure}{S\arabic{figure}}%
\renewcommand{\thetable}{S\arabic{table}}%

\begin{center}
{\large \textbf{Supplementary Material}}
\end{center}

This supplementary material consists of the detailed analytical derivation of Eq.~(6) of the main text as well as additional numerical results.

\section{Analytics -- Supplementary}

To describe the current-induced magnetization dynamics of the considered ferromagnetic thin film we use the Landau-Lifshitz-Gilbert equation for the unit vector field $\vect{M}$ {generalized to include} spin-torque effects due to the electric current:
\begin{equation}
(\partial_{t} + v_{s} \partial_{x}) \hat{\vect{M}} = -\gamma \hat{\vect{M}} \times \vect{B}_{\mathrm{eff}} + \alpha \hat{\vect{M}} \times \biggl( \partial_{t} + \frac{\beta}{\alpha} v_{s} \partial_{x} \biggr) \hat{\vect{M}},
\end{equation}
where $\gamma$ is the gyromagnetic ratio, and $\alpha$ and $\beta$ are {the} dimensionless Gilbert {damping} and non-adiabatic {spin-transfer-torque} parameters.
The effective magnetic field is given by $\vect{B}_{\mathrm{eff} }= -M_{s}^{-1} (\delta F[\hat{\vect{M}}]/\delta \hat{\vect{M}})$, where $F[\hat{\vect{M}}]$ describes the free energy of the system and $M_{s}$ is the saturation magnetization.
The applied uniform DC current along the $x$ direction enters the equation via the effective spin velocity: $v_{s} = \xi j$ with $\xi = g P \mu_{B}/(2 e M_{s})$, where $g$ is the g-factor, $P$ is the current polarization, $\mu_{B}$ is the Bohr magneton, $e$ is the electron charge, and $j$ is the current.

In the following we assume the absence of non-adiabatic spin-torque, \textit{i.e.}, $\beta = 0$, as it is typically small.
However, we checked numerically that including the non-adiabatic spin-torque term does not change qualitatively the results.
We will therefore use in the following
\begin{equation}
\label{eq:LLG}
(\partial_{t} + \xi j \partial_{x}) \hat{\vect{M}} = \gamma \hat{\vect{M}} \times \frac{\delta F[\hat{\vect{M}}]}{\delta \hat{\vect{M}}} + \alpha \hat{\vect{M}} \times \partial_{t} \hat{\vect{M}}.
\end{equation}

In the following we will first describe the setup.
Then we will show that a small dissipation allows one to construct a functional which must be minimized by a static solution in the presence of a current.
Using this functional we demonstrate that above a certain current $j_{c}$ the static solution is not stable.
By finding the approximate form of the static magnetization configuration for the current just below $j_{c}$, we derive the dynamics of the magnetization for the homogeneous DC currents just above $j_{c}$ and prove the creation of periodic magnetic textures.

\subsection{Pinning setup as boundary value problem}

To describe the periodic texture formation analytically we assume the following general setup, that includes the more precise setup we have proposed in the main text, shown in Fig.~1.
We consider a two-dimensional magnetic film where the corresponding model, \textit{i.e.}, the magnetic free energy $F[\hat{\vect{M}}(\vect{r})]$ contains in particular exchange interactions and a uniaxial anisotropy along a certain direction given by $\vect{e}_{a}$ and does not explicitly depend on time.
Furthermore there is a small region in the magnetic film where the direction of the uniaxial anisotropy is modified.

As in the main text, we will call this region of modified anisotropy a pinning center.
The direction along which the magnetization in the center of this region is pinned is along $\vect{e}_{p}$.
Without loss of generality we assume that the pinning center is located at the origin of our coordinate system.
We consider these two constraints as boundary conditions on the vector field $\hat{\vect{M}}(\vect{r})$ constituting the pinning problem:
\begin{subequations}
\begin{align}
\label{eq:boundaryA}
\hat{\vect{M}}(\vect{r} \to \infty, t) &= \vect{e}_{a}, \\
\label{eq:boundaryP}
\hat{\vect{M}}(\vect{r} = 0, t) &= \vect{e}_{p}.
\end{align}
\end{subequations}
Note that the fact that we consider the pinning constraint as a boundary condition also allows to consider the magnetic free energy functional $F[\hat{\vect{M}}(\vect{r})]$ of the system to be translationally invariant along the $x$ direction.

As explained in the main text, when increasing the current strength, first the magnetic texture around the pinning center will deform and elongate until it finally rips off from the pinning center, eventually forms a skrymion and antiskyrmion pair that then travels along the magnetic film.

\subsection{Energy considerations}

Rewriting Eq.~\eqref{eq:LLG} as
\begin{equation}
\label{eq:FpOdot}
\partial_{t} (\gamma F + j \Omega) = -\alpha \int \bigl( \partial_{t} \hat{\vect{M}} \bigr)^{2} d^{2}r,
\end{equation}
where
\begin{equation}
\partial_{t} \Omega \equiv \xi \int \hat{\vect{M}} \cdot (\partial_{x} \hat{\vect{M}} \times \partial_{t}\hat{\vect{M}}) d^{2}r,
\end{equation}
allows to interpret the change of the magnetic free energy $F$ in time by two sources: i) dissipation due to Gilbert damping and ii) work done by the current.
The power supplied by the current $-j \partial_{t} \Omega$ is linear in the applied current density $j$ and describes the ``effective electric field'' generated by the magnetization dynamics.

To obtain a deeper understanding of the functional $\Omega$ it is instructive to consider the boundary value problem for the magnetization configuration $\hat{\vect{M}}(\vect{r}, t) \equiv \hat{\vect{M}}(x, y, t)$ at a fixed $y$ component, and thereby mapping the problem for each $y$ onto an effective one-dimensional model.
According to Eq.~\eqref{eq:boundaryA} $\hat{\vect{M}}(x \to -\infty, y, t) = \hat{\vect{M}}(x \to \infty, y, t) = \vect{e}_{a}$ allows to identify the one-dimensional model as a closed line on the unit sphere with a corresponding $y$-dependent solid angle $\omega_{y}$.
The change of the solid angle $\omega_{y}$ over time $dt$ is given by $d\omega_{y} = dt \int \hat{\vect{M}} \cdot (\partial_{x} \hat{\vect{M}} \times \partial_{t} \hat{\vect{M}}) dx$.
By comparing the definition of $\partial_{t} \Omega$ in Eq.~\eqref{eq:FpOdot} with the expression of the solid angle, we see that the functional $\Omega$ can be interpreted as the sum of the solid angles over all $y$ components:
\begin{equation}
\Omega = \xi \int \omega_{\vect{y}} dy.
\end{equation}
As the dissipative coefficient $\alpha$ is positive, the value of the functional
\begin{equation}
\label{eq:action}
\mathcal{S}[\hat{\vect{M}}(\vect{r})] = \gamma F + j \Omega
\end{equation}
appearing in Eq.~\eqref{eq:FpOdot} is decreasing during the time evolution.
In particular, on the static solution the value of this functional must be at its minimum while satisfying the non-linear constraint $|\hat{\vect{M}} (\vect{r})| = 1$.

\subsection{Instability and critical current}

Let's consider a static solution $\hat{\vect{M}}_{0}(\vect{r})$ for some particular current strength $j_{0}$.
To analyse its stability we expand around the solution and consider $\hat{\vect{M}}(\vect{r}) = \hat{\vect{M}}_{0}(\vect{r}) + \hat{\vect{m}}(\vect{r})$, where $|\hat{\vect{m}}| \ll 1$ and $\hat{\vect{M}}(\vect{r})$ still satisfies the (normalized) boundary value problem.
This implies for the vector field $\hat{\vect{m}}$ the following boundary conditions:
\begin{equation}\label{eq:linearBoundary}
\hat{\vect{M}}_{0}(\vect{r}) \perp \hat{\vect{m}}, \qquad \hat{\vect{m}}(\vect{r} \to \infty) \to 0, \qquad \hat{\vect{m}}(\vect{r} = 0) = 0,
\end{equation}
where the last two conditions are the consequences of Eqs.~\eqref{eq:boundaryA} and \eqref{eq:boundaryP}.
A static solution $\hat{\vect{M}}_{0}(\vect{r})$ for some particular current strength $j_{0}$ is stable, if the operator $\hat{\Pi}_{0}$ defined by
\begin{equation}
\hat{\Pi}_{0} \hat{\vect{m}}(\vect{r}) \equiv \int d^2r' \frac{\delta^{2} \mathcal{S}_{0}}{\delta \hat{\vect{M}}(\vect{r}) \delta \hat{\vect{M}}(\vect{r'})} \biggr|_{\hat{\vect{M}} = \hat{\vect{M}}_{0}(\vect{r})} \hat{\vect{m}}(\vect{r'})
\end{equation}
is positive definite, \textit{i.e.}, has only strictly positive eigenvalues, for all functions satisfying the conditions defined by Eqs.~\eqref{eq:linearBoundary}.
Note that here the index ``$0$'' means that the functional $\mathcal{S}$ is evaluated at the current density $j_{0}$.

Additionally we know that both the Hamiltonian and $\Omega$ are translationally invariant.
Thus, if a static solution $\hat{\vect{M}}_{0}(\vect{r})$ at current strength $j_{0}$ is an extremum of the functional $\mathcal{S}$ of Eq.~\eqref{eq:action}, then a solution $\hat{\vect{M}}_{0}(\vect{r} + a \vect{e}_{x})$ shifted by some distance $a$ is also an extremum of the functional $\mathcal{S}$ with the same extremal value for any distance $a$.
Taking the limit $a \to 0$ we find that
\begin{equation}
\hat{\Pi}_{0} \partial_{x} \hat{\vect{M}}_{0}(\vect{r}) = \int d^2r' \frac{\delta^{2} \mathcal{S}_{0}}{\delta \hat{\vect{M}}(\vect{r}) \delta \hat{\vect{M}}(\vect{r'})} \biggr|_{\hat{\vect{M}} = \hat{\vect{M}}_{0}(\vect{r})} \partial_{x} \hat{\vect{M}}_{0}(\vect{r'}) = 0,
\end{equation}
meaning that the field $\partial_{x} \hat{\vect{M}}_{0}(\vect{r})$ is always a zero mode of our linear operator $\hat{\Pi}_{0}$.
Note, however, that the function $\partial_{x} \hat{\vect{M}}_{0}(\vect{r})$ is not in the defined functional space, as for arbitrary currents it does not satisfy the last of the conditions given by Eq.~\eqref{eq:linearBoundary}.
However, if there exists a current $j_{c}$ such that the corresponding magnetization profile $\hat{\vect{M}}_{c}(\vect{r})$ fulfills also the last condition
\begin{equation}
\label{eq:jcritical}
\partial_{x} \hat{\vect{M}}_{c}(\vect{r} = 0) = 0,
\end{equation}
then one of the eigenvalues of the operator $\hat{\Pi}_{c}$ is zero and the solution becomes unstable.
Eq.~\eqref{eq:jcritical} defines implicitly the value of the critical current $j_{c}$ above which the solution is no longer stable.

We note that our construction of the functional $\mathcal{S}$, as well as the derivation of the critical current $j_{c}$, relies on the fact that at infinity the magnetization is uniform.
It is known, however, that at currents larger than a certain current $j_{c*}$, such a uniform state becomes unstable.\cite{Bazaliy98, Shibata2005, Sitte2016}.
Thus, our derivation makes sense only if $j_{c} < j_{c*}$ which is the case as long as the pinning direction $\vect{e}_{p}$ does not coincide with the direction of the uniaxial anisotropy $\vect{e}_{a}$.

To further understand analytically the picture of the instability where at the critical current density $j_{c}$ the magnetic textures change from static to dynamic, we consider in the following only currents close the critical one.
For such currents the magnetization is either static or changes with time very slowly.
More crucially, the statement of either static or very slow magnetization changes allows to also neglect the Gilbert damping term in the analytic calculations, provided that magnetization dynamics is deduced from the physical correct static magnetization profile.
Notably it is this damping term that ensures that all spin wave solutions are decaying with time and then allows to extract the physically correct static magnetization out of numerous static solutions of Eq.~\eqref{eq:LLG}.
In contrast this term is essential for the numerical solutions and all simulations were performed including an Gilbert damping term.

\subsection{Static solutions for currents just below the critical current}

Within this subsection we derive the relation of how much a magnetic texture is shifted by an electrical current strength just below the critical one.

As was shown above, at the critical current strength $j_{c}$ the operator $\hat{\Pi}_{j_{c}}$ has a zero mode.
For the currents just below the critical current this zero mode acquires a small gap.
So, for the currents just below $j_{c}$ this mode can be taken into account explicitly.
Hence, the magnetization configuration at a certain current $j_{0} \lesssim j_{c}$ is just the configuration $\hat{\vect{M}}_{c}(\vect{r})$ -- the static solution at the critical current -- shifted by some small distance $x_{0}$ along the direction of the current plus a small correction: $\hat{\vect{M}}_{0}(\vect{r}) \equiv \hat{\vect{M}}_{c}(\vect{r} - x_{0} \vect{e}_{x}) + \delta \hat{\vect{M}}$.
Note that the distance $x_{0}$ is characteristic for the applied current strength $j_{0}$.
The condition of satisfying the two boundary conditions $\hat{\vect{M}}_{0}(0) = \hat{\vect{M}}_{c}(0)$, \textit{i.e.}, $\delta \hat{\vect{M}}(0) = 0$, and $\hat{\vect{M}}_{0}^{2}(\vect{r}) = 1$ up to order $x_{0}^{2}$ leads to:
\begin{subequations}
\begin{align}
\hat{\vect{M}}_{0}(\vect{r})
&= \hat{\vect{M}}_{c}(\vect{r} - x_{0} \vect{e}_{x}) - \frac{1}{2} x_{0}^{2} \partial_{x}^{2} \hat{\vect{M}}_{c}(\vect{r}) - \frac{1}{2} x_{0}^{2} \hat{\vect{M}}_{c}(\vect{r}) (\partial_{x} \hat{\vect{M}}_{c})^{2} + \mathcal{O}(x_{0}^{2}) \\
\label{eq:Sjb}
&= \hat{\vect{M}}_{c}(\vect{r}) - x_{0} \partial_{x} \hat{\vect{M}}_{c}(\vect{r}) - \frac{1}{2} x_{0}^{2} \hat{\vect{M}}_{c}(\vect{r}) (\partial_{x} \hat{\vect{M}}_{c})^{2} + \mathcal{O}(x_{0}^{2}).
\end{align}
\end{subequations}
Thus, for each small $x_{0}$ we have a specific current density $j_{0}$ for which the above function is a solution up to the order $x_{0}^{2}$.
Let's now exploit the fact that $\hat{\vect{M}}_{0}(\vect{r})$ is a static solution of the boundary value problem and thus minimizes the functional $\mathcal{S}$ given by Eq.~\eqref{eq:action}, which for above ansatz yields
\begin{align*}
\mathcal{S}_{0}[\hat{\vect{M}}_{0}(\vect{r})]
&= \mathcal{S}_{c}[\hat{\vect{M}}_{0}(\vect{r})] + (j_{0} - j_{c}) \Omega[\hat{\vect{M}}_{0}] \\
&= \mathcal{S}_{c}[\hat{\vect{M}}_{c}(\vect{r})] + (j_{0} - j_{c}) \Omega[\hat{\vect{M}}_{c}] - x_{0} \int d^2r \frac{\delta \mathcal{S}_{c}}{\delta \hat{\vect{M}}(\vect{r})} \biggr|_{\hat{\vect{M}}_{c}(\vect{r})}  \partial_{x} \hat{\vect{M}}_{c}(\vect{r}) \\
&\quad + \frac{x_{0}^{2}}{2} \int d^2r d^2r' \frac{\delta^{2} \mathcal{S}_{c}}{\delta \hat{\vect{M}}(\vect{r}) \delta \hat{\vect{M}}(\vect{r}')} \biggr|_{\hat{\vect{M}}_{c} (\vect{r})} \partial_{x} \hat{\vect{M}}_{c}(\vect{r}) \partial_{x} \hat{\vect{M}}_{c}(\vect{r}') \\
&\quad - \frac{1}{2} x_{0}^{2} \int d^2r \frac{\delta \mathcal{S}_{c}}{\delta \hat{\vect{M}}(\vect{r})} \biggr|_{\hat{\vect{M}}_{c}(\vect{r})} \hat{\vect{M}}_{c}(\vect{r}) (\partial_{x} \hat{\vect{M}}_{c}(\vect{r}))^{2} + \mathcal{O}(x_{0}^{2}).
\end{align*}
The first term on the RHS is independent of the distance $x_{0}$.
The third and the fourth term vanish, as i) $\partial_{x} \hat{\vect{M}}_{c}(\vect{r})$ satisfies Eq.~\eqref{eq:linearBoundary} and the operator $\frac{\delta \mathcal{S}_{j_{c}}}{\delta \hat{\vect{M}}(\vect{r})}$ is zero on all fields satisfying the conditions \eqref{eq:linearBoundary}, and ii) $\partial_{x} \hat{\vect{M}}_{c}(\vect{r})$ is per definition the zero mode of the operator $\hat{\Pi}_{c}$.
The last term is the interesting one.
It does not vanish as the field $\hat{\vect{M}}_{c}(\vect{r}) (\partial_{x} \hat{\vect{M}}_{c}(\vect{r}))^{2}$ does not satisfy the conditions~\eqref{eq:linearBoundary} -- in fact $\frac{\delta \mathcal{S}_{c}}{\delta \hat{\vect{M}} (\vect{r})} \parallel \hat{\vect{M}}_{c}(\vect{r})$.
So finally we get:
\begin{equation}
\label{eq:S0M0}
\mathcal{S}_{0}[\hat{\vect{M}}_{0}(\vect{r})] \approx \mathcal{S}_{c}[\hat{\vect{M}}_{c}(\vect{r})] + [\mathcal{A}_{c} x_{0}^{2} + (j_{0} - j_{c})] \Omega_{c}[\hat{\vect{M}}_{c}],
\end{equation}
where
\begin{equation}
\mathcal{A}_{c} \equiv -\frac{1}{2 \Omega_{c}[\hat{\vect{M}}_{c}]} \int d^2r \frac{\delta \mathcal{S}_{c}}{\delta \hat{\vect{M}}(\vect{r})} \biggr|_{\hat{\vect{M}}_{c}(\vect{r})} \hat{\vect{M}}_{c}(\vect{r}) (\partial_{x} \hat{\vect{M}}_{c}(\vect{r}))^{2}.
\end{equation}
As $\mathcal{S}_{c}[\hat{\vect{M}}_{c}(\vect{r})]$, $\Omega_{c}[\hat{\vect{M}}_{c}(\vect{r})]$, and $\mathcal{A}_{c}$ are independent of the distance $x_{0}$, minimizing Eq.~\eqref{eq:S0M0} with respect to $x_{0}$ while satisfying the condition that for $x_{0} = 0$ the critical current $j_{c}$ and $j_{0}$ coincide, then yields:
\begin{equation}
\label{eq:jb}
j_{c} - j_{0} = x_{0}^{2} \mathcal{A}_{c}.
\end{equation}
This equation establishes the relation between the shift of the critical solution $x_{0}$ and the current $j_{0}$.
Note that the difference $j_{c} - j_{0}$ is indeed second order in $x_{0}$.

\subsection{Periodic texture production for currents just above the critical current}

For currents just larger than the critical current $j_{c}$, defined by exhibiting a zero mode of the operator $\hat{\Pi}_{c}$, the static solutions are unstable.
At currents $j$ just above $j_{c}$ this zero mode becomes dynamic.
Therefore, to obtain information about the dynamics of the magnetization configuration, we make the following ansatz for the applied current $j$:
\begin{equation}
\hat{\vect{M}}(\vect{r}, t) = \hat{\vect{M}}_{0}(\vect{r}, t) + \hat{\vect{m}}(\vect{r}, t),
\label{eq:ansatzT}
\end{equation}
where $\hat{\vect{M}}_{0}(\vect{r}, t)$ is the static magnetization configuration for the current $j_{0}$, given by Eq.~\eqref{eq:jb} with time-dependent $x_{0}$.
The dynamics of the zero mode means that we consider $x_{0}$ to be time-dependent.
$j_{0}(t)$ satisfies Eq.~\eqref{eq:jb} at each moment of time.
Furthermore, $\hat{\vect{m}}(\vect{r}, t)$ is a small perturbation obeying the following conditions:
\begin{equation}
\hat{\vect{M}}_{0}(\vect{r}, t) \perp \hat{\vect{m}}(\vect{r}, t), \qquad \hat{\vect{m}}(r \to \infty, t) \to 0, \qquad \hat{\vect{m}}(r = 0, t) = 0.
\label{eq:boundaryT}
\end{equation}
The above ansatz we insert into the LLG equation Eq.~\eqref{eq:LLG}.
Linearizing Eq.~\eqref{eq:LLG} in $\vect{m}(r,t)$ we obtain for $\alpha = 0$:
\begin{multline}
\label{eq:dyn0}
\frac{\partial \hat{\vect{M}}_{0}(\vect{r})}{\partial t} + \partial_{t} \hat{\vect{m}} + \xi j \partial_{x} \hat{\vect{M}}_{0}(\vect{r}) \\
= \gamma \hat{\vect{M}}_{0}(\vect{r}) \times \int d^2r' \frac{\delta^{2} F[\hat{\vect{M}}]}{\delta \hat{\vect{M}}(\vect{r}) \delta \hat{\vect{M}}(\vect{r'})} \biggr|_{\vect{M}_{0}(\vect{r})} \hat{\vect{m}}(\vect{r}', t) + \hat{\vect{m}} \times \frac{\delta F[\hat{\vect{M}}(\vect{r})]}{\delta \hat{\vect{M}}(\vect{r})} \biggr|_{\hat{\vect{M}}_{0}(\vect{r})}.
\end{multline}
Using further $\mathcal{S}[\hat{\vect{M}}_{0}] = \mathcal{S}_{0}[\hat{\vect{M}}_{0}] + (j - j_{0}) \Omega[\hat{\vect{M}}_{0}]$ and the definition of $\Omega[\hat{\vect{M}}_{0}]$ we obtain
\begin{multline}
\label{eq:dyn}
\partial_{t} \hat{\vect{m}} - \int d^2r' \hat{\vect{M}}_{0}(\vect{r}) \times \frac{\delta^{2} \mathcal{S}_{0}}{\delta \hat{\vect{M}}(\vect{r}) \delta \hat{\vect{M}}(\vect{r'})} \biggr|_{\hat{\vect{M}}_{0}(\vect{r})} \hat{\vect{m}}(\vect{r}', t) - \hat{\vect{m}} \times \frac{\delta \mathcal{S}_{0}}{\delta \hat{\vect{M}}(\vect{r})} \biggr|_{\hat{\vect{M}}_{0}(\vect{r})} \\
= -\frac{\partial \hat{\vect{M}}_{0}(\vect{r})}{\partial{t}} - (j - j_{0}) \partial_{x} \hat{\vect{M}}_{0}(\vect{r}) = \partial_{x} \hat{\vect{M}}_{c} [\partial_{t} {x}_{0} - (j - j_{0})] + \mathcal{O} (x_{0}^3).
\end{multline}
In the last equation we have used Eq.~\eqref{eq:Sjb} and the fact that $\partial_{t} x_{0}$ and the current difference is of the order of $\partial_{t} x_{0} \sim (j - j_{0}) \sim x_{0}^{2}$.
Equation \eqref{eq:dyn} is an inhomogeneous linear equation for the vector field $\hat{\vect{m}}(\vect{r}, t)$ which must satisfy all the conditions in Eq.~\eqref{eq:boundaryT}.
For
\begin{equation}
\label{eq:bDynamics}
\partial_{t} x_{0} = j - j_{0}
\end{equation}
Eq.~\eqref{eq:dyn} has a simple trivial solution, $\hat{\vect{m}} \equiv 0$, satisfying all conditions given by Eqs.~\eqref{eq:boundaryT}.
Equation~\eqref{eq:bDynamics} together with Eq.~\eqref{eq:jb} gives the dynamics of the soft mode of the magnetization configuration for the currents just above $j_{c}$.
To find the period of the texture production we need to solve Eqs.~\eqref{eq:bDynamics} and \eqref{eq:jb}: $\partial_{t} x_{0} = j - j_{c} + \mathcal{A}_{c} x_{0}^{2}$, or $t = \int \frac{dx}{j - j_{c} + \mathcal{A}_{c} x^{2}}$.
The major contribution to the integral comes from $\mathcal{A}_{c} x^{2} < j - j_{c}$, and the integral converges fast for larger $x$.
In the texture production period calculation we can extend the integration from $+\infty $ to $-\infty$, leading to
\begin{equation}
\label{eq:period}
T = \int^{\infty }_{-\infty} \frac{dx}{j - j_{c} + \mathcal{A}_{c} x^{2}} = \frac{\pi}{\sqrt{\mathcal{A}_{c}}} \frac{1}{\sqrt{j - j_{c}}}.
\end{equation}
This is the period of production of the textures in magnetization configuration.

\section{Numerics -- Supplementary}

In the following we provide more simulation results.
In particular we consider the theoretical case without \emph{any} ``twisting'' interaction, \textit{i.e.}, also without dipolar interactions.
Then we provide examples where different magnetic textures are shedded.
At the end we show that in principle one can use this set-up to generate skyrmion/antiskyrmion pairs in a pulse-operation mode.
Finally, we provide an overview of parameters used in our simulations in Table~\ref{tab:SimulationParameters} including parameters of Refs.~\citenum{Heinonen2016, Jiang2016} for reference.

\subsection{Without dipole-dipole interactions}

\begin{figure}[tbp]
\centering
\includegraphics[width=.65\columnwidth]{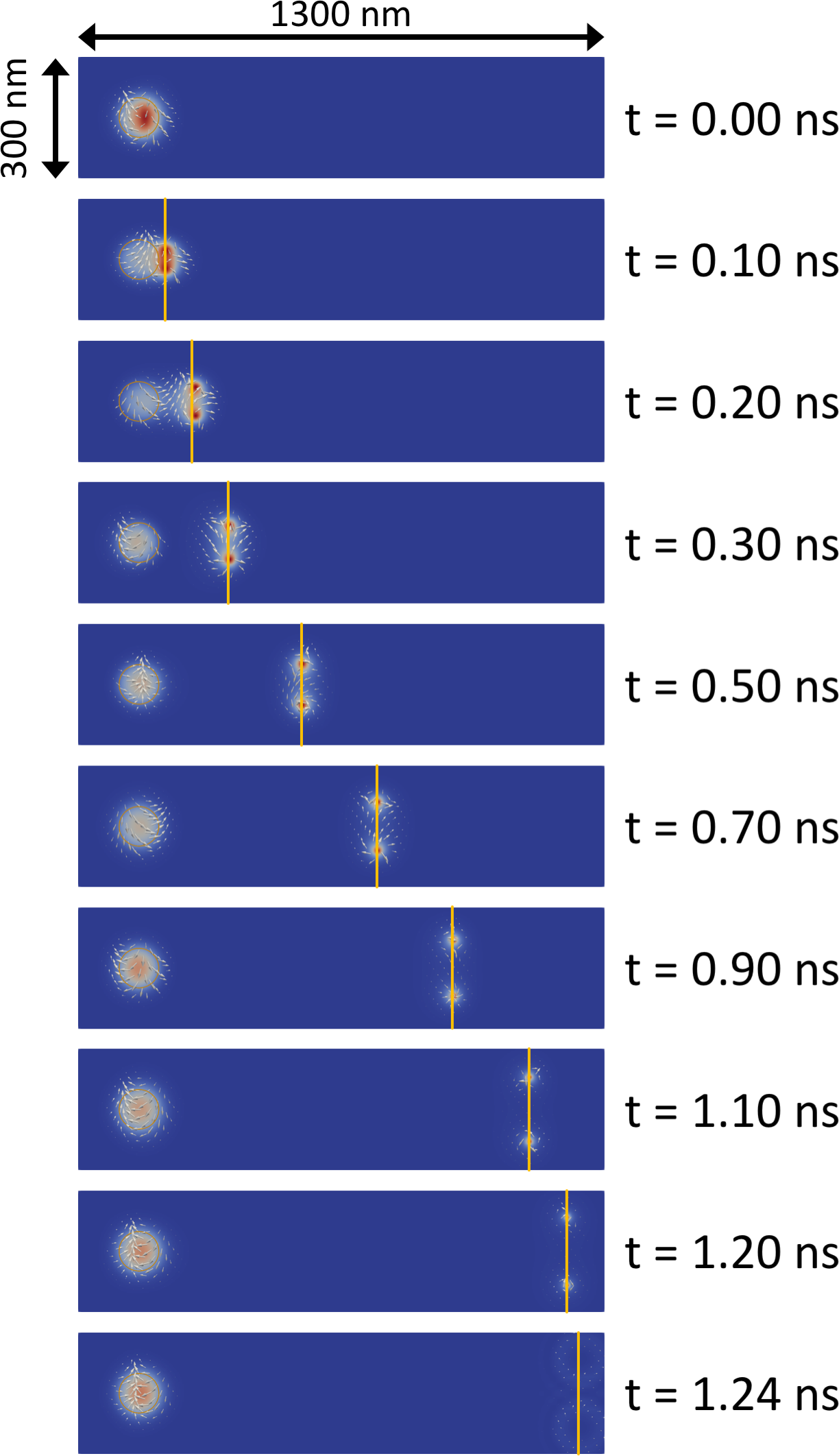}
\caption{%
Shedding of skyrmion/antiskyrmion pairs without dipolar interactions.
For this setup, we used an effective uniaxial anisotropy $z$ direction to incorporate the effects of the dipolar field: $M_{s} = \SI{3e5}{\ampere\per\meter}$, $A_{\mathrm{ex}} = \SI{2e-11}{\joule\per\meter}$, $K_{u} = \SI{2.345e4}{\joule\per\cubic\meter}$, $K_{u}^{p} = \SI{-5.655e4}{\joule\per\cubic\meter}$ and $\tilde{K}_{u}^{p} = \SI{1e3}{\joule\per\cubic\meter}$ \emph{inside} the pinning area which has a radius of \SI{50}{\nano\meter}, and Gilbert damping of $\alpha = 0.05$, and $\beta = 0$.
Applied current density is $\SI{5e12}{\ampere\per\square\meter}$.
}
\label{fig:without_dipolar_field}
\end{figure}

As can be deduced from analytics no "twisting" interaction is needed for the shedding process itself, \textit{i.e.}, also in theoretical version without dipolar interactions a shedding process still occurs.
This we have also confirmed numerically, see Fig.~\ref{fig:without_dipolar_field}.
Note that in the absence of any twisting interactions including dipolar fields the skyrmion and the antiskyrmion configuration are energetically degenerate and therefore they also decay at the same time.

\subsection{Shedding of different magnetic textures}

\begin{figure}[tbp]
\centering
\includegraphics[width=.5\columnwidth]{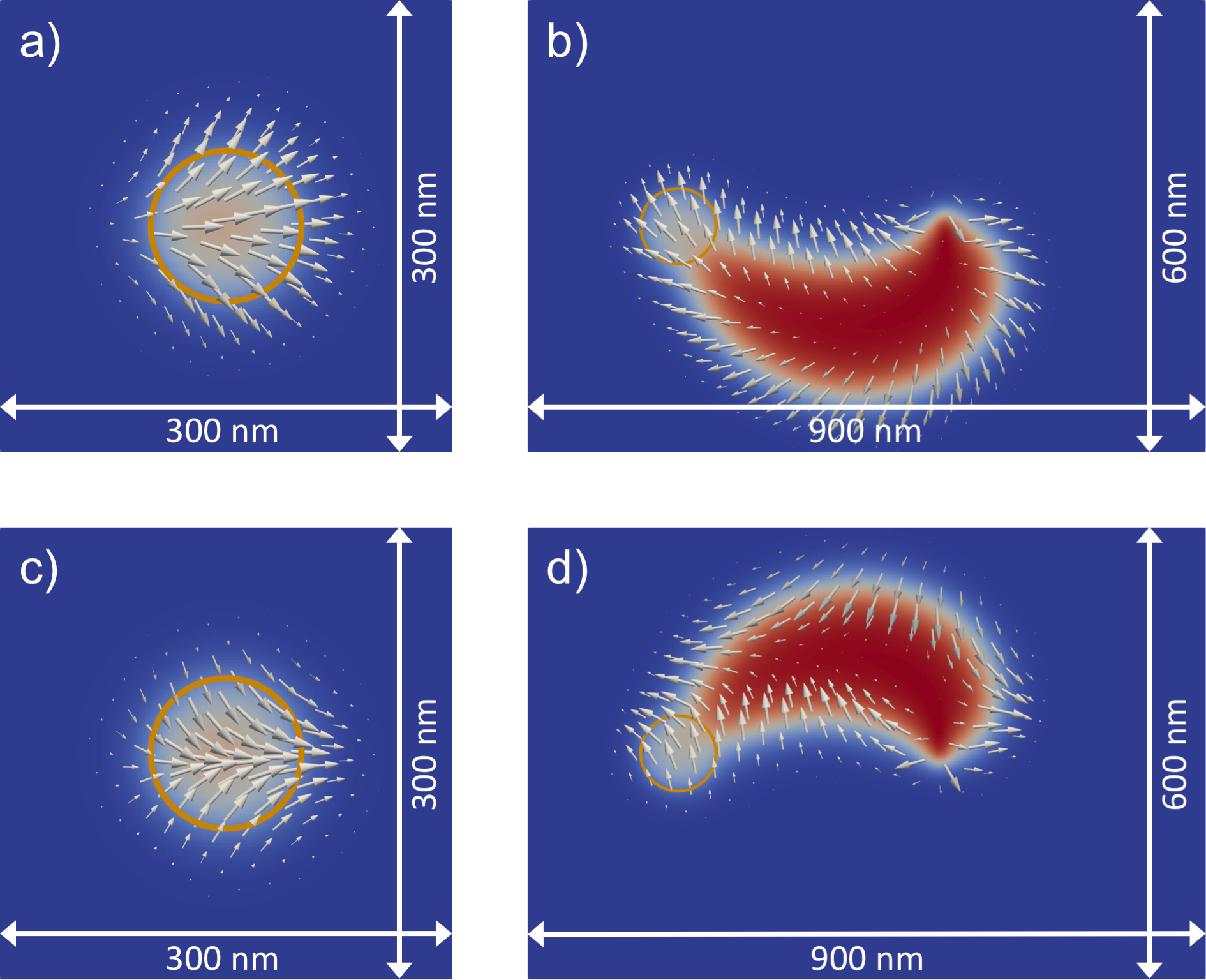}
\caption{%
Simluation results for the creation process in the presence of larger (anisotropic) DMI.
The arrows visualize the in-plane component of the magnetization.
The yellow circle indicates the defect region with modified anisotropy.
The top (bottom) panel shows the results for $D_1 = \pm D_2= \SI{1e-3}{\joule\per\square\meter}$.
The left column shows the starting configuration for the $z$-component around the pinning center whereas the right column shows the time evolution corresponding to the DC current density $j = \SI{3e12}{\ampere\per\square\meter}$.
The other parameters used for this setup are: $M_{s} = \SI{3e5}{\ampere\per\meter}$, $A_{\mathrm{ex}} = \SI{2e-11}{\joule\per\meter}$, $K_{u} = \SI{8e4}{\joule\per\cubic\meter}$, $K_{u}^{p} = 0$ and $\tilde{K}_{u}^{p} = \SI{1e3}{\joule\per\cubic\meter}$ \emph{inside} the pinning area which has a radius of \SI{50}{\nano\meter}, and Gilbert damping of $\alpha = 0.05$, and $\beta = 0$.
}
\label{fig:largeDMI}
\end{figure}

\begin{figure}[tbp]
\centering
\includegraphics[width=.99\columnwidth]{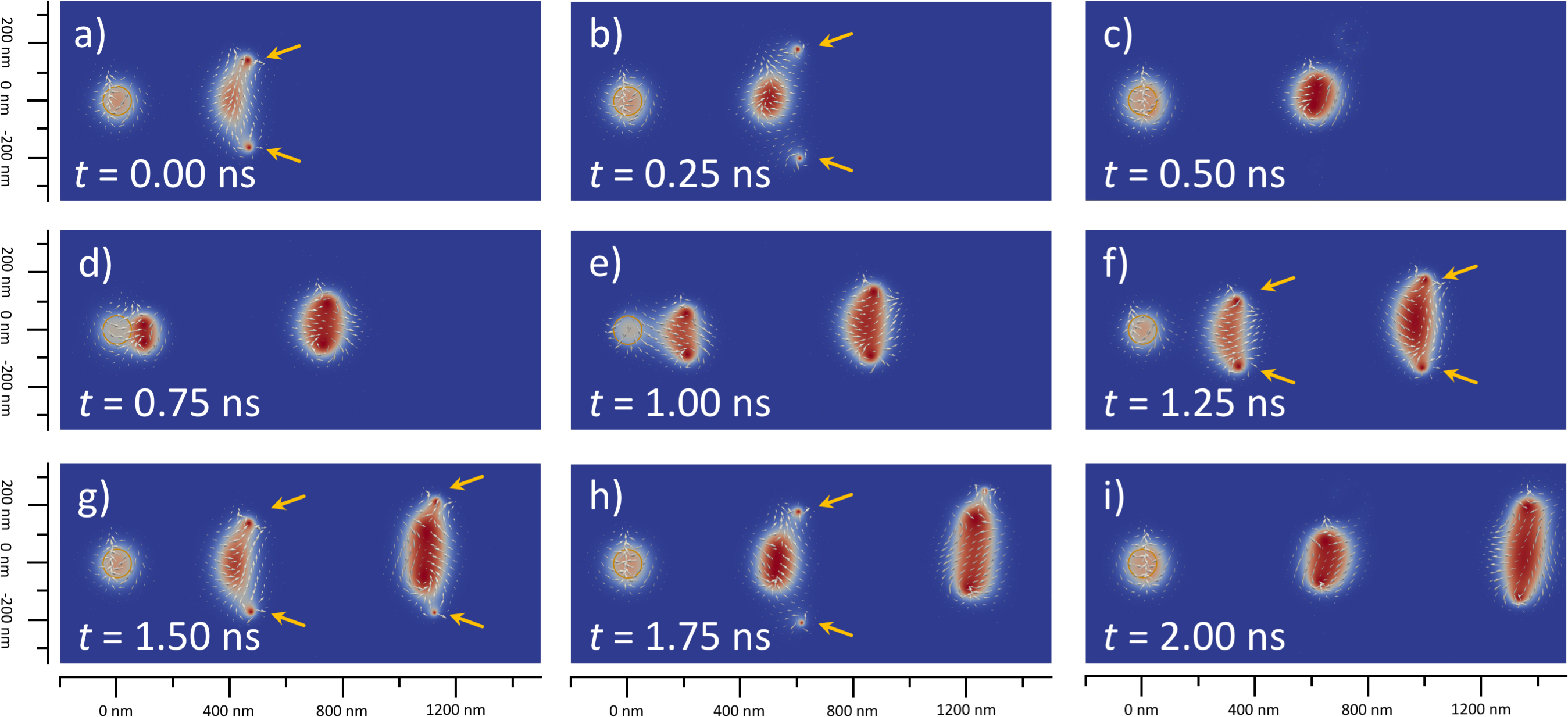}
\caption{%
Shown here are more complicated shedded magnetic textures.
The parameters used for this simulation are: $M_{s} = \SI{6.5e5}{\ampere\per\meter}$, $A_{\mathrm{ex}} = \SI{2e-11}{\joule\per\meter}$, $K_{u} = \SI{2.88e5}{\joule\per\cubic\meter}$, $K_{u}^{p} = \SI{1e5}{\joule\per\cubic\meter}$ and $\tilde{K}_{u}^{p} = \SI{1e3}{\joule\per\cubic\meter}$ \emph{inside} the pinning area which has a radius of \SI{50}{\nano\meter}, and Gilbert damping of $\alpha = 0.25$, and $\beta = 0$.
Applied current density is \SI{7e12}{\ampere\per\square\meter}.
}
\label{fig:largeMS}
\end{figure}

Depending on the material parameters, the current strength and the size of the pinning center also different magnetic textures can be shedded.
Here we show two examples of different shedded magnetic textures.
Compared to the main text we have chosen i) a system with a larger DMI, see Fig.~\ref{fig:largeDMI}, where the center of the shedded magnetic textures is not along the direction of the current but determined by the interplay of current and DMI, and ii) a setup with a larger saturation magnetization, see Fig.~\ref{fig:largeMS}.
Here we observe, aside from the skyrmion/antiskyrmion pairs, also the shedding of other magnetic textures that look like ``walls'' from which then on the edges also skyrmion and antiskyrmions rip off.

\subsection{Pulse-operation mode}

\begin{figure}[tbp]
\includegraphics[width=.9\columnwidth]{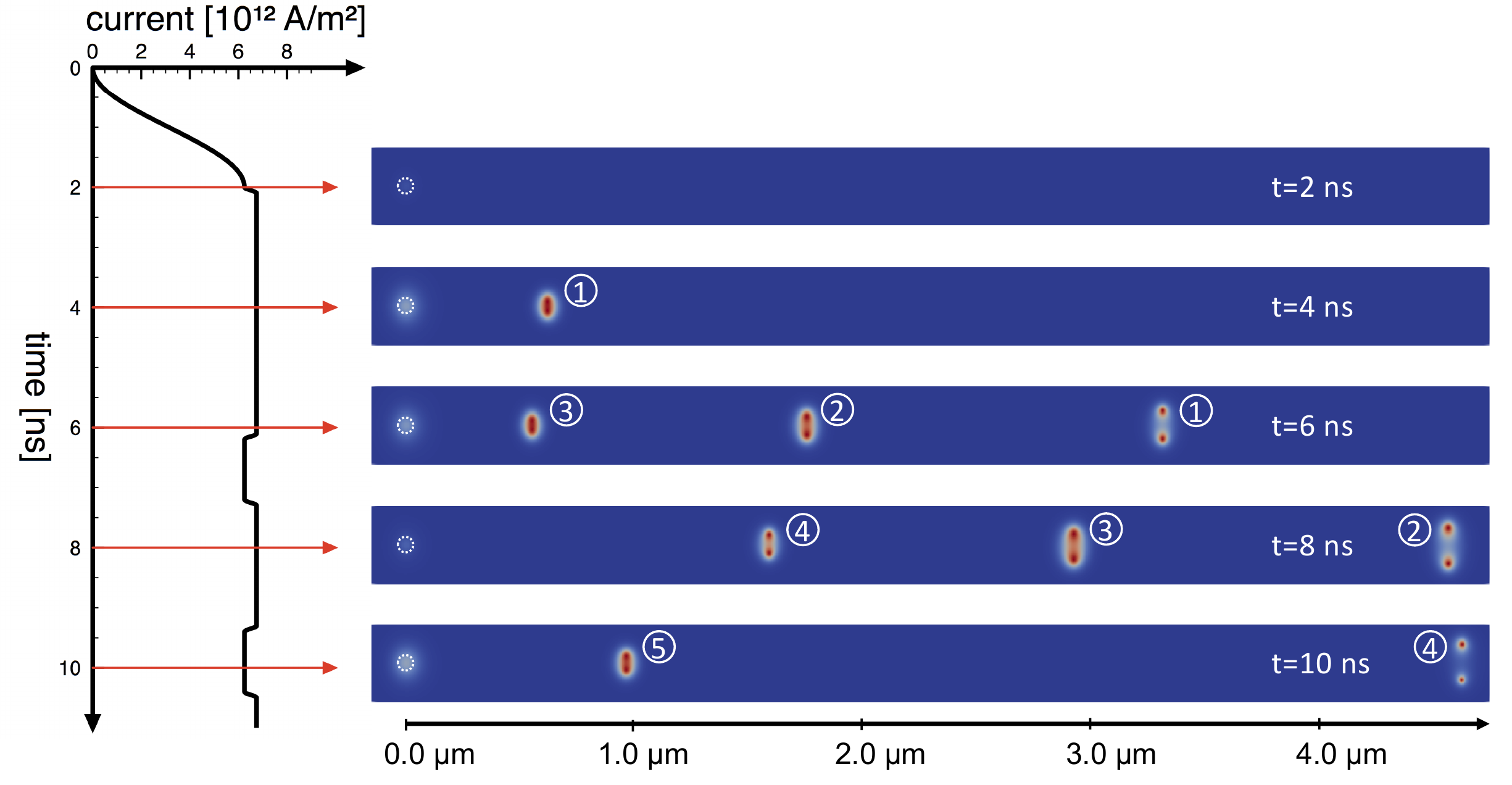}
\caption{
Skyrmion creation by DC pulse operation mode.
A desired number of skyrmions can be created by controlling the strength of the applied DC current.
As shown in the last pulse, also a single skyrmion-anti-skyrmion pair can be created.
To show the clean mechanism of the pulse-operation mode, in these simulations we have chosen a setup that reduces the effect spin waves.
The particular parameters that we have chosen are: $M_{s} = \SI{1.33e5}{\ampere\per\meter}$, $A_{\mathrm{ex}} = \SI{1.3e-11}{\joule\per\meter}$, $K_{u} = \SI{2e4}{\joule\per\cubic\meter}$, and $K_{u}^{p} = 0$ \emph{inside} the pinning area which has a radius of \SI{25}{\nano\meter}, and Gilbert damping of $\alpha = 0.01$, and $\beta = 0$.
Pinning of the magnetization in $x$ direction is achieved via a local magnetic field of $H = \SI{5e4}{\ampere\per\meter}$.
Applied current density is switched between \SI{6.25e12}{\ampere\per\square\meter} and \SI{6.75e12}{\ampere\per\square\meter}.
}
\label{fig:pulse_mode}
\end{figure}

Executing the device in a pulse operation mode, in principle allows to create a desired number of skyrmions with any specific spacing pattern.
A simulation result for the pulse operation in time is shown in Fig.~\ref{fig:pulse_mode}.
In this simulation we have created a pinning center by adding a local magnetic field instead of modifying the anisotropy strength and we have used the current pulse sequence shown in the left part of Fig.~\ref{fig:pulse_mode}.
Switching on the DC current, the magnetic texture around the pinning center first elongates until it breaks off and then starts shedding skyrmion/antiskyrmion pairs until the shedding rate quickly reaches its nominal rate.
A similar transient state occurs when lowering the DC current below $j_{c}$ where the pinning center keeps shedding skyrmion/antiskyrmion pairs until it has lost enough ``momentum'' and shedding stops.
The duration of the transient states depends on the details of the system.
An important aspect of this mode is the lower power that it requires.
The creation process requires a substantially higher current density than the one to move the magnetic textures.
Hence, by shortening the pulses that create the skyrmion-anti-skyrmion pairs and using a much lower current density to move them, one can minimize Joule heating.

\begin{table}[tbp]
\centering
\begin{tabular}{l|cccccccc|cc}
\toprule
\textbf{parameters used in Fig.}                       & 2, 4, 5, 6 & 7a         & 7b         & S1$^{\dagger}$ & S2a,b      & S2c,d       & S3         & S4$^\ddagger$ & Ref.~\citenum{Heinonen2016} & Ref.~\citenum{Jiang2016} \\
\hline
$A_{\mathrm{ex}}$ [\si{\pico\joule\per\meter}]         & \num{20}   & \num{20}   & \num{20}   & \num{20}       & \num{20}   & \num{20}    & \num{20}   & \num{13}      & \num{30}                    & \num{10}   \\
$M_{s}$ [\si{\kilo\ampere\per\meter}]                  & \num{300}  & \num{300}  & \num{300}  & \num{300}      & \num{300}  & \num{300}   & \num{650}  & \num{133}     & \num{650}                   & \num{650}  \\
$K_{u}$ [\si{\kilo\joule\per\cubic\meter}]             & \num{80}   & \num{80}   & \num{80}   & \num{23.45}    & \num{80}   & \num{80}    & \num{288}  & \num{20}      & \num{288}                   & \num{288}  \\
$K_{u}^{p}$ [\si{\joule\per\cubic\meter}]              & \num{0}    & \num{0}    & \num{0}    & \num{-56.66}   & \num{0}    & \num{0}     & \num{100}  & \num{0}       & ---                         & ---        \\
$\tilde{K}_{u}^{p}$ [\si{\kilo\joule\per\cubic\meter}] & \num{1}    & \num{1}    & \num{1}    & \num{1}        & \num{1}    & \num{1}     & \num{1}    & ---           & ---                         & ---        \\
$j$ [\si{\mega\ampere\per\square\centi\meter}]         & \num{500}  & \num{325}  & \num{325}  & \num{500}      & \num{300}  & \num{300}   & \num{700}  & \num{675}     & \num{10}                    & \num{1.3}  \\
$\alpha$                                               & \num{0.05} & \num{0.05} & \num{0.05} & \num{0.05}     & \num{0.05} & \num{0.05}  & \num{0.25} & \num{0.01}    & \num{0.02}                  & \num{0.02} \\
$P$                                                    & \num{1.0}  & \num{1.0}  & \num{1.0}  & \num{1.0}      & \num{1.0}  & \num{1.0}   & \num{1.0}  & \num{0.56}    & ---                         & ---        \\
pinning center radius [\si{\nano\meter}]               & \num{50}   & \num{50}   & \num{50}   & \num{50}       & \num{50}   & \num{50}    & \num{50}   & \num{25}      & ---                         & ---        \\
$D_{1}$ [\si{\milli\joule\per\square\meter}]           & \num{0}    & \num{0.5}  & \num{0.5}  & \num{0}        & \num{1.0}  & \num{1.0}   & \num{0}    & \num{0}       & \num{0.5}                   & \num{0.5}  \\
$D_{2}$ [\si{\milli\joule\per\square\meter}]           & \num{0}    & \num{0.5}  & \num{-0.5} & \num{0}        & \num{1.0}  & \num{-1.0}  & \num{0}    & \num{0}       & \num{0.5}                   & \num{0.5}  \\
\botrule
\end{tabular}
\caption{%
Overview over simulation parameters used in the main text and supplementary material.
The parameters used for the figures in the main text as well as Figs.~\ref{fig:without_dipolar_field} and \ref{fig:largeDMI} are motivated by Refs.~\citenum{Zheng2002, Roy2003, Navas2010, Kane2015}.
In all simulations we have set $\beta = 0$.
$^{\dagger}$~In Fig.~\ref{fig:without_dipolar_field} we have incorporated the effect of the dipolar field into renormalized effective uniaxial anisotropies along the $z$ direction.
$^{\ddagger}$~In Fig.~\ref{fig:pulse_mode} we have used a local magnetic field of \SI{50}{\kilo\ampere\per\meter} instead of an in-plane symmetry-breaking anisotropy field to tilt the magnetization into the $x$ direction inside the pinning area.
Here the parameters are similar to Ref.~\citenum{Sitte2016}.
}
\label{tab:SimulationParameters}
\end{table}

\end{document}